\documentclass{aa}
\usepackage{graphicx}
\usepackage[varg]{txfonts}
\usepackage{natbib}
\usepackage{array}

\begin{document}

\title{Mid-infrared blends and continuum signatures of dust drift and accretion in protoplanetary disks}

\author{Antonellini, S.$^{1,2,3}$\and Kamp, I.$^{1}$\and Waters, L.B.F.M.$^{4,5}$} 
\authorrunning{Antonellini et al.}

\institute{Kapteyn Astronomical Institute, Postbus 800, 9700 AV Groningen, The Netherlands\\
\email{antonellini@astro.rug.nl}\and 
Astrophysics Research Centre, School of Mathematics and Physics, Queens University Belfast, Belfast BT7 1NN, UK\and 
Institut de RadioAstronomie Millim\' etrique (IRAM), 300 Rue de la Piscine, Domaine Unversitaire, 38400, St Martin D'Heres, France\
 \and
Department of Astrophysics/IMAPP, Radboud University, PO Box 9010, 6500 GL Nijmegen, the Netherlands
\and
SRON Netherlands Institute for Space Research, Niels Bohrweg 4, 2333 CA Leiden, the Netherlands
}

\abstract
 {The mid-infrared (MIR) emission of molecules such as H$_2$O, HCN, OH, CO$_2$, and C$_2$H$_2$, has been identified in the Spitzer Infrared
Spectrograph (IRS) spectra of many protoplanetary disks. According to the modelling results, the blend strengths are affected by different disk properties such as the gas mass and dust content in the disks. An observational correlation between HCN and water blend fluxes has been noted, specifically related to a changing disk gas mass.}
{We aim to find out whether the explanation for the observed flux correlation between HCN and water in the MIR could also be attributed to other properties and processes taking place in disks, such as the evolution of dust grains. We also consider what the consequences of these results would be in relation to the disk evolution.}
 {We used pre-existing ProDiMo radiation thermal-chemical disk models exploring a range of properties such as the disk gas mass, disk inner radius, dust size power law distribution, and, finally, time-dependent dust evolution. From these models, we computed the MIR fluxes of HCN and H$_2$O blends. Simultaneously, we derived the spectral indices from the simulated spectral energy distributions (SEDs) in the Spitzer IRS regime. Finally, we compared these quantities with the observed data.}
 {The MIR blend fluxes correlation between HCN and water can be explained as a consequence of dust evolution, namely,\ changes in the dust MIR opacity. Other disk properties, such as the disk inner radius and the disk flaring angle, can only partially cover the dynamic range of the HCN and water blend observations. At the same time, the dynamic range of the MIR SED slopes is better reproduced by the disk structure (e.g. inner radius, flaring) than by the dust evolution. Our model series do not reproduce the observed trend between continuum flux at 850~$\mu$m and the MIR HCN/H$_2$O blend ratio. However, our models show that this continuum flux is not a unique indicator of disk mass and  it should therefore be used jointly with complementary observational data for optimal results.} 
 {The presence of an anti-correlation between MIR H$_2$O blend fluxes and the MIR SED is consistent with a scenario where dust evolves in disks, producing lower opacity and stronger features in the Spitzer spectral regime, while the gas eventually becomes depleted at a later stage, leaving behind an inner cavity in the disk.} 


\maketitle

\section{Introduction}\label{sec:intro}\vspace{5mm}

Protoplanetary disks have been extensively observed in the mid-infrared (MIR) regime \citep[e.g.][]{pontoppidan1,pontoppidan2,salyk,salyk2,banz12}, which is the spectral regime that is spatially related to the terrestrial-planet formation region. Many molecules have been observed in the Spitzer Infrared Spectrograph (IRS) spectral range \citep[$\sim\!\!5\!\!-\!\!35~\mu$m;][]{houck}, including OH and H$_2$O, along with organics such as HCN and C$_2$H$_2$ \citep[e.g.][]{pontoppidan,zhang,salyk1,salyk2}. Previous studies investigated the observed existence of a correlation between the blend fluxes from MIR H$_2$O blends and those produced by organic molecules in the same regime, using slab models \citep{carr,carr1,najita1}; these works showed that the HCN/H$_2$O blend ratio in the MIR is correlated with the disk gas mass (as derived from the dust mass by scaling with a constant gas-to-dust mass ratio). From the observations, MIR molecular blend {\bf{fluxes}} (in particular H$_2$O) are found to be anti-correlated to the dust outer radius as measured from the continuum emission \citep{bp2020}. An anti-correlation also exists between MIR H$_2$O blend {\bf{fluxes}} and the measured spectral index $n_{13-30}$, as seen in both \citet{salyk1,bp2020}. Both of these works have suggested a plausible link between dust and gas that is potentially due to icy pebble migration across the snow line.

Modelling of the 1D radial drift of icy pebbles together with chemistry \citep{bosman} showed that the gas phase CO$_2$ abundance gets enhanced by more than three orders of magnitude if sublimation from radially inward drifting icy grains is included. Using the DALI model, these authors showed that this produces a CO$_2$ band strength, which is beyond what is consistent with current observations, and this requires a revision of the processes that would lead to a destruction of the molecule. \citet{greenwood} coupled the two-pop-py code for dust growth and radial drift \citep{birnstiel} with ProDiMo thermo-chemical disk models \citep[e.g.][]{woitke} to show that dust grain evolution changes the dust opacity in the inner disk; even by simply allowing this form of opacity change, the CO$_2$ MIR blend fluxes are shown to increase with time. So, the increase in blend fluxes can also be achieved without taking ice transport and sublimation into account (i.e.\ without a change in element abundances). Other MIR-emitting molecules such as HCN, NH$_3$, H$_2$O, and OH, are affected in the same way by an evolving dust opacity and their blend fluxes get up to 2~dex stronger on a timescale of 10~Myr \citep{greenwood1}.

In this work, we investigate whether grain size and opacity changes due dust evolution, namely, growth, radial drift, and vertical settling, have the capacity explain the observed H$_2$O versus HCN correlation in disks around T~Tauri stars as well as the absolute  blend fluxes. With the use of previously published advanced radiation thermo-chemical models \citep{antonellini,greenwood}, we investigate the blend fluxes of H$_2$O and HCN in the MIR regime (Sect.~\ref{sec:mod}). To corroborate our findings, we then discuss the effects of dust evolution on the disks' spectral energy distribution (SED) in the MIR and (sub)mm wavelengths as well as on the 10 and 20~$\mu$m silicate features. We compare these results with observed trends in Sect.~\ref{sec:cmp}. Our interpretation and conclusions are presented in Sects.~\ref{sec:dis} and \ref{sec:con}. 

\section{Modeling}\label{sec:mod}\vspace{5mm}

We made use of the ProDiMo thermo-chemical disk modelling code \citep{woitke, kamp4, thi2011}. We also used the T~Tauri disk models previously published by \citet{antonellini,antonellini1} and by \citet{greenwood}. The first modelling series is composed of steady-state chemistry models (\citet{antonellini,antonellini1}), varying model parameters such as disk gas mass ($M_\mathrm{gas}$, with $M_\mathrm{dust}$ fixed), dust size power law distribution ($a_\mathrm{pow}$), flaring index ($\beta$), and inner radius ($R_\mathrm{in}$).  An overview is presented in Table~\ref{One} and Appendix~\ref{Ap1}. These model series include dust settling according to the prescription detailed in \citet{dubrulle}. All these models have been re-run for this paper, including the non-LTE ro-vibrational cooling for HCN, OH, C$_2$H$_2$, CH$_4$, NH$_3$, and CO$_2$, as described in \citet{woitke6}. 
This allowed us to obtain blend vertical escape probability fluxes for all these molecules and to calculate MIR blend emission spectra.

\begin{table}
\caption{Disk model series used in this study}
\centering
\begin{tabular}{lp{2.5cm}p{2.5cm}}\hline\hline
Model series & \citet{antonellini} & \citet{greenwood}\\ \hline
M$_\mathrm{gas}$ [M$_{\odot}$] & 10$^{-5}$-10$^{-1}$ & 0.1 (initial)$^{(*)}$\\ 
a$_\mathrm{pow}$ [-] & 2.0-4.5 & evolving\\
$\beta$ [-] & 0.8-1.25 & 1.15 \\
R$_\mathrm{in}$ [au] & 0.05-55 & 0.07\\
\hline
\end{tabular}
\tablefoot{$^{(*)}$ Due to viscous accretion there is a  drop in the gas mass during the disk time evolution, which is up to 50\% of the initial value.}
\label{One}
\end{table}

A second model series considered in our analysis includes the effects of time-dependent radial drift, growth, and fragmentation of dust grains in a typical T~Tauri disk with ages ranging from 20~kyr up to 10~Myr \citep{greenwood}. 
The models couple the dust evolution code two-pop-py from \citep{birnstiel2012} with the thermo-chemical code ProDiMo. 
The evolution model series starts from a more massive disk ($M_{\rm disk}$=0.1$M_\odot$), which is smaller ($R_{\rm taper}$=100~au), less turbulent ($\alpha_{\rm vis}$=10$^{-3}$), and  also contains  larger dust grains ($a_{\rm max}$=190~cm), with respect to the standard one considered in the parameterised steady state model series. In addition, these models compute  the dust evolution purely (e.g.\ transport, dust settling), without taking ice sublimation at the respective ice line locations into account; thus, they ignore any potential gas enrichment in the inner disk. More details of the model parameters used for this T~Tauri disk model can be found in \citet[][Table~1]{greenwood}.

From the comparison between the parameterised steady state models and the time evolution series at earlier epochs ($t\,\le\,5.6\,\times\,10^{5}$~yr), we find the blend flux predictions and continuum slopes to be in the same range of values ($10^{-19}$-$10^{-16}$~W/m$^2$, -0.5 to 0.5, respectively). This demonstrates that the differences in fundamental disk parameters outlined above do not affect the combined study of all these models. 

\subsection{Blend fluxes}\label{sub:fl}

From our radiation thermo-chemical models, we obtained vertical escape probability fluxes of H$_2$O and HCN in the 10-40~$\mu$m regime. Fluxes obtained with this method are based on a rotational LTE treatment of HCN, which is reliable in the MIR regime, given the emitting region conditions, as discussed in \citet{woitke6}. A non-LTE study of HCN by \citet{Bruderer2015} found that blend fluxes are very close to LTE values and only the blend profiles change; at the spectral resolution of Spitzer ($R\,\sim\,600$), this will thus not affect our results. 

For comparison with observations, we coadded the fluxes of all the blends used by \citet{najita1} in the 14-17~$\mu$m regime, obtaining fluxes for the HCN blend at 14~$\mu$m (integrated in the range 13.905-14.049~$\mu$m) and the three H$_2$O blends at 17.12~$\mu$m (integrated in the range 17.077-17.155~$\mu$m), 17.22~$\mu$m (integrated in the range 17.155-17.300~$\mu$m), and 17.36~$\mu$m (integrated in the range 17.300-17.480~$\mu$m). 
This procedure is equivalent to that applied to the observational data (continuum subtraction and integration of blend flux).
In the case of HCN, we excluded from the co-adding all H$_2$O and OH blends included in the spectral range, in order to be consistent with the method applied to the observational data.
To correct for the effect of blending, \citet{najita1} fitted the strong H$_2$O blends in the 15~$\mu$m regime and then subtracted the slab model fit from the MIR spectrum in the HCN blend wavelength range. 
In our case, we got the individual blend fluxes contributing to each blend from the model. We then simply sum up the fluxes of all the blends contributing to each blend in the same spectral ranges as indicated above. This procedure is equivalent to the method of \citet{najita1} applied to the IRS observed spectra.

\subsection{Continuum}\label{sub:co}

To investigate how individual disk properties affect the continuum, \citet{salyk2} and \citet{najita1} used the spectral slope in the MIR wavelength range.
The spectral index (as a proxy for the SED slope) is defined in general as \citep[see][]{furlan1}:
\begin{equation}\label{eq2}
    n_\mathrm{a-b}=\frac{\log_{10}F_\mathrm{b}\lambda_\mathrm{b}-\log_{10}F_\mathrm{a}\lambda_\mathrm{a}}{\log_{10}\lambda_\mathrm{b}-\log_{10}\lambda_\mathrm{a}}\,\,\, ,
\end{equation}
where $F_\mathrm{a}$, and $F_\mathrm{b}$ are the fluxes in Jy at the wavelengths $\lambda_\mathrm{a}$ and $\lambda_\mathrm{b}$, respectively. The two observational data sets used in our study applied different values of $\lambda_b$, that is, 25 and 30 $\mu$m, respectively.
To enlarge the overlap between available MIR blend fluxes and continuum slope measurements, we decided to include the spectral indices defined in the range 13-30 and 13-25~$\mu$m.
In our full sample we have 3 targets overlapping, that we can use to assess the deviation between the two spectral index measurements. We find then that the average difference between these spectral indexes 13-30 and 13-25~$\mu$m is $\sim$0.7, such a difference would eventually move the data from \citet{salyk2} to lower spectral indexes, thereby covering a region of the plot already described by some of our models (Fig.~\ref{fig:model}). Thus, the data from the two works can be combined. 

\section{Comparisons with observations}\label{sec:cmp}\vspace{5mm}

In the following, we test the hypothesis that dust evolution (growth, destruction, and radial migration, and thus with no transport of ices) alone can explain the observed H$_2$O and HCN blend fluxes, and the observed correlation between their blend ratio and disk mass in T~Tauri disks. We do this by comparing the model series predictions for molecular blend fluxes and SED spectral indices with the observations. Comparing a broader range of observables can break degeneracies that could exist when only comparing fluxes or flux ratios.

\subsection{MIR-blend fluxes}

\begin{figure}
\includegraphics[width=0.5\textwidth]{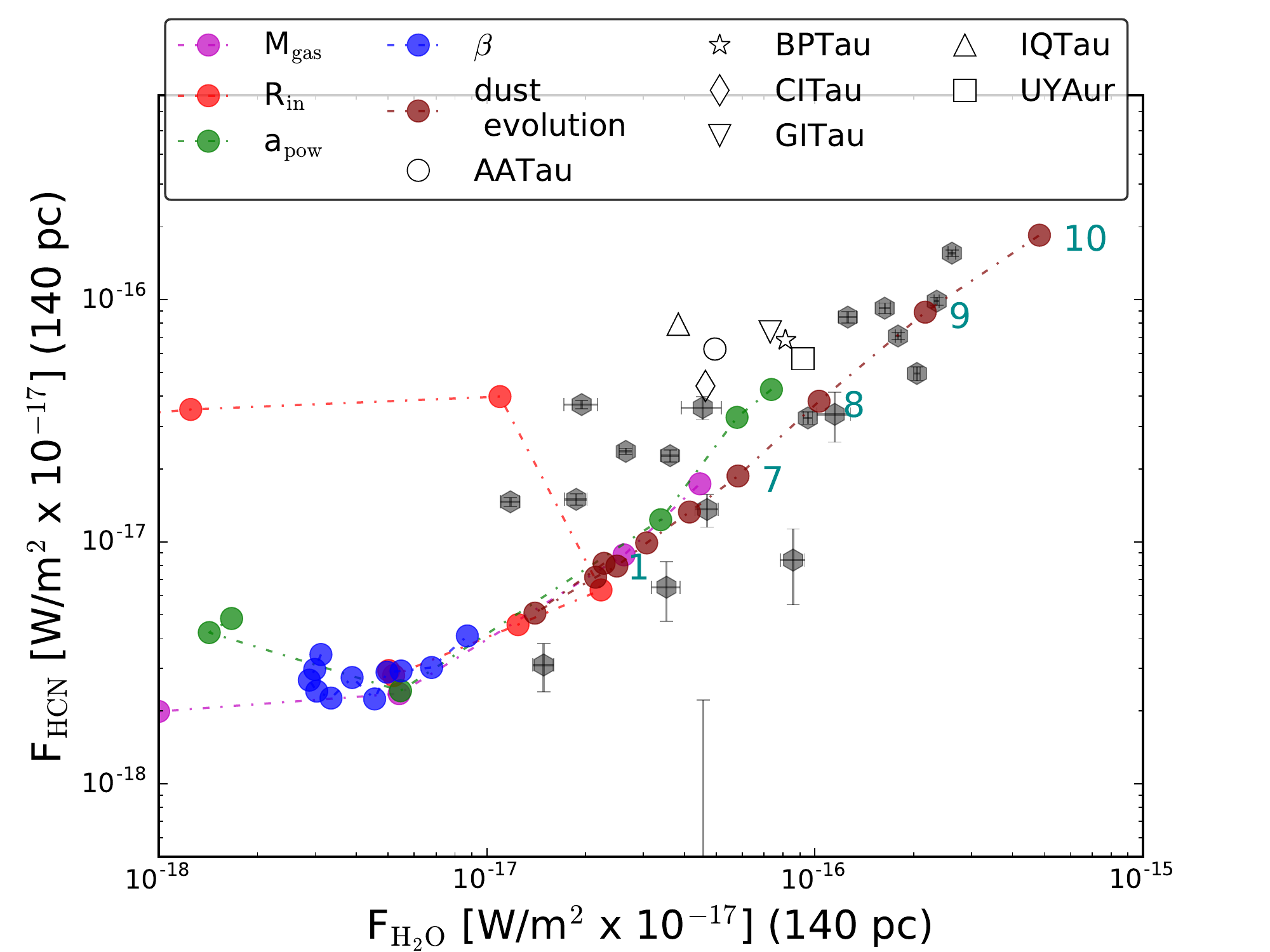}
\caption{Observed HCN versus H$_2$O MIR blend fluxes from \citet{najita1} with error bars (grey hexagons), scaled to a distance of 140~pc using GAIA parallaxes\protect\footnote{https://gea.esac.esa.int/archive/}. The model series for disk gas mass (magenta circles), disk inner radius (red circles), dust power law index (green circles), disk flaring index (blue circles) from \citet{antonellini,antonellini1}, and dust evolution from \citet{greenwood1} (brown circles) are over-plotted. Some of the dust evolution models are labelled in dark cyan, with the number indicating the model time step. Observed objects deviating from the trend of the dust evolution model series are plotted as open symbols and are named in the legend.}
\label{fig:model}
\end{figure}

The model blend fluxes extracted from the two modelling series for HCN and H$_2$O are compared to the observations from \citet{najita1} in Fig.~\ref{fig:model}. Our steady-state models with different physical disk parameters and homogeneous dust properties are not able to reach the observed high values of both HCN and H$_2$O MIR blend fluxes ($>\!10^{-16}~$W/m$^2$). 
Models with different gas mass (and constant $M_\mathrm{dust}$), $M_\mathrm{gas}$, flaring angle, $\beta$, and power law index of grain size distribution, $a_\mathrm{pow}$, show blend fluxes that reach also below the Spitzer detection limits ($<\!10^{-17}~$W/m$^2$ for HCN and $<\!10^{-18}~$W/m$^2$ for H$_2$O). Previous modelling works have also often found overly weak MIR molecular blends and it was then necessary to impose a gas-to-dust ratio much higher than 100 in order to match observed blend fluxes \citep[e.g.][]{Meijerink, Bruderer2015, woitke6}.
The only two simulations that reach high blend fluxes are models with an inner radius larger than the standard one of 0.1~au (here the HCN blend flux increases by more than a dex) and the dust size power-law distribution, which is part of the process included in the dust evolution model series. The HCN ro-vibrational transitions are calculated using LTE for the rotational population levels. This likely over-predicts the blend fluxes in disk regions where LTE does not hold, as in disks with an inner hole, and the emission of the molecule is produced in more optically thin layers \citep{woitke6}. \citet{Bruderer2015} used 2D disk models (DALI code) and a detailed HCN model molecule to show that the difference between LTE and non-LTE blend fluxes for HCN remains within a factor of 3.

Models including dust evolution (grain growth, fragmentation, and radial migration) can cover the full observed dynamic range. They can explain H$_2$O blend fluxes above 10$^{-17}$~W/m$^2$ and the water blend flux versus the continuum anti-correlation trend shown in \citet[][see Fig.~\ref{Acc}]{bp2020}. This model series matches well the observed trend that has been reported by \citet{najita1} and follow an almost linear relation between $F_\mathrm{HCN}$ and $F_\mathrm{H_2O}$. The reason behind it is a reduced continuum opacity in the MIR-emitting regions of HCN and H$_2$O; both are located inside of 2~au \citep{greenwood1}. This disk region from which the blends are emitted, is characterised by high densities, high temperatures and a high gas-to-dust ratio increasing with time: the gas density is $>\!10^{8}$~cm$^{-3}$ and increasing up to 2 dex between 10$^{4}$ and 10$^{7}$~yr (Fig.~\ref{EmReg}), the g/d ratio evolves from a few 1000 to $\sim10^5$ (at $\sim 10$~Myr), while the gas temperature stays roughly constant between $300$ and $400$~K for both H$_2$O and HCN \citep{greenwood1}.
Under these conditions the chemistry reaches steady-state on timescales of  between 10$^2$-10$^3$~yr.

\subsection{Mid-IR H$_2$O and the continuum slope}
\label{sec:continuum-slope}

\begin{figure}
\includegraphics[width=0.5\textwidth]{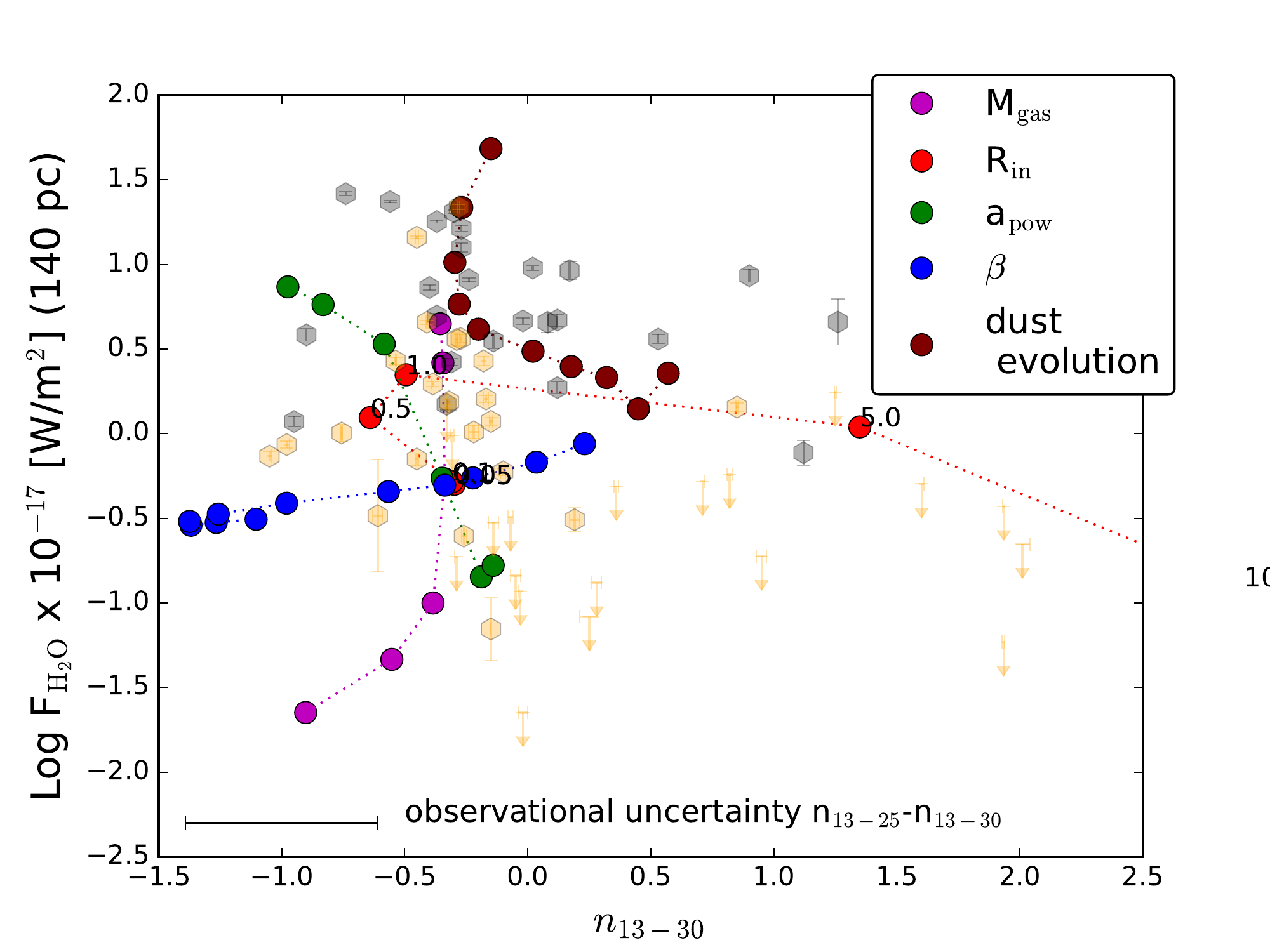}
\caption{H$_2$O MIR fluxes versus $n_{13-30}$ from \citet{salyk2} (orange hexagons) and $n_{13-25}$ from \citet{najita1} (grey hexagons) scaled to 140~pc\protect\footnote{Distances are taken from https://gea.esac.esa.int/archive/.} with error bars in grey.  The model series for disk gas mass (purple circles), disk inner radius (red circles), dust power law index (green circles), disk flaring index (blue circles) from \citet{antonellini,antonellini1}, and dust evolution from \citet{greenwood} (0.018~Myr, 0.032~Myr, 0.056~Myr, 0.1~Myr, 0.18~Myr, 0.32~Myr, 0.56~Myr, 1.0~Myr, 1.8~Myr, 3.2~Myr, brown circles) are over-plotted. Bottom error bar shows the average difference detected between the three targets for which continuum indexes at 13-30 and 13-25 were available.}\label{due}
\label{fig:model1}
\end{figure}

Previous analysis of our model series already showed an intrinsic connection between MIR blend fluxes and the neighbouring continuum, where each disk parameter variation creates a unique effect on MIR H$_2$O blends and on the MIR continuum \citep[e.g.][ also see the introduction]{antonellini,antonellini1,antonellini2}. Having found above that the dust evolution series can reproduce the observed strength of HCN and water fluxes as well as their correlation, we want to test this further by checking whether or not the same model series also reproduces the water versus SED slope plane of observations. We used the MIR blend fluxes and continuum spectral indexes from Table 4 of \citet{salyk2} and Table 1 of \citet{najita1} for the same objects. We compared these with the respective observables derived from our models in the same way (see Sects.~\ref{sub:fl} and~\ref{sub:co}).
While the observed trend in HCN and water molecular blend fluxes is naturally explained by dust evolution in the inner disk, the same disk model series can only reproduce part of the observed range of MIR spectral indices. There is not a simple correlation between MIR water and continuum spectral indexes (Fig.~\ref{due}). \citet{salyk2} interpreted the trend with spectral index as disks with a bluer continuum being associated with more settled dust grains. 
More quantitatively, \citet{najita1} defined targets with a spectral index less than 0.5 as normal T~Tauri disks, while targets with spectral indices larger than 1.5 are defined as transition disks. While our dust evolution series does include settling, it does not capture the opening of gaps and holes associated with transition disks. For spectral indices of $<$-0.5, our model predictions cover the same range of values in the water blend flux versus spectral index plane as the observations.

 
The flaring index (growing), the gas mass, and the dust power law exponent can explain a redder MIR SED (negative indices, $n_{13-25}\!<\!-0.5$). The intermediate region (-0.5$\!<\!n_{13-25}\!<\!0.5$) may represent disks with small gaps, such that the objects cannot yet clearly being classified as transition disks \citep[][]{evans}. This interpretation is supported by the number of MIR H$_2$O blend detections. Some potential transition disks have detected MIR H$_2$O, which could be consistent with a disk where dust is evolving and a clearing-out process is happening, two strong indicators of ongoing planet formation.

Observations with a blue MIR SED ($n_{13-25}\!>\!0.5$) are  covered by models with an inner cavity larger than 15~au. According to \citet{furlan2}, transition disks have $n_{13-25}\!>\!1.5$. In  Fig.~\ref{fig:model1}, we included the sample from \citet{salyk2}, which contains many upper limits for H$_2$O MIR blend fluxes. 

Only the inner radius model series extends to very high MIR SED indexes ($n_{13-30}\!>\!1$ from 5~au). These models also predict very low and undetectable MIR H$_2$O blend fluxes (till 10$^{-20}$~W/m$^2$), largely in agreement with observations. Very negative spectral index values are typically found for debris disks \citep{furlan2}. 

On the red SED side, our dust evolution model series cannot reach spectral indexes lower than -0.5. On the contrary, our flaring model, the pure dust size power law distribution and the gas mass allow to reach this blue region of the spectral index. 

In \citet{antonellini}, we already showed that increasing the dust-to-gas ratio, increasing the turbulent mixing coefficient, or steepening the power law dust size distribution (separate aspects of dust evolution), all produce a reddening in the MIR SED. This is consistent with what we now report here for the more self-consistent dust evolution model series (brown filled circles in Fig.~\ref{fig:model1}). 
On the other side, the presence of objects with a spectral index less than -0.5 \citep[the so called 'anemic' disks;][]{lada2,evans1} may require a different explanation. The dust evolution model used in \citet{greenwood} is based on a set of parameters such as $\alpha_\mathrm{vis}$, dust sizes, initial distribution, and dust-to-gas mass ratio; also, it does not cover the full range of possible scenarios we may encounter in reality. We can also not claim that the set of adopted parameters are representative of all disks in this observational sample. Along this line of thought, our dust evolution series is able to cover an upper region of the plot in Fig.~\ref{fig:model1}, but it also merely covers one aspect of disk evolution, namely, the one related to dust. If this were combined with some other evolution aspect of disks, such as flaring angle and mass dispersion, this could cover the dynamic range described by the observations even at lower H$_2$O fluxes and lower MIR indexes. 

\subsection{Disk mass and the HCN/H$_2$O blend ratio}

\citet{najita1} found a trend between the HCN/H$_2$O blend ratio and the disk mass derived from sub-mm data. They took the disk gas mass either derived from SEDs using power law disk models or estimated from the 850~$\mu$m continuum flux using a simple power law \citep{andrews2}. 


We extracted the flux at $850$~$\mu$m from our model series. We note that in our modelling series, the disk dust mass changes only in the dust evolution series (brown); all other model series assume a constant disk dust mass of $10^{-4}$~M$_\odot$. 

\begin{figure}
\includegraphics[width=0.5\textwidth]{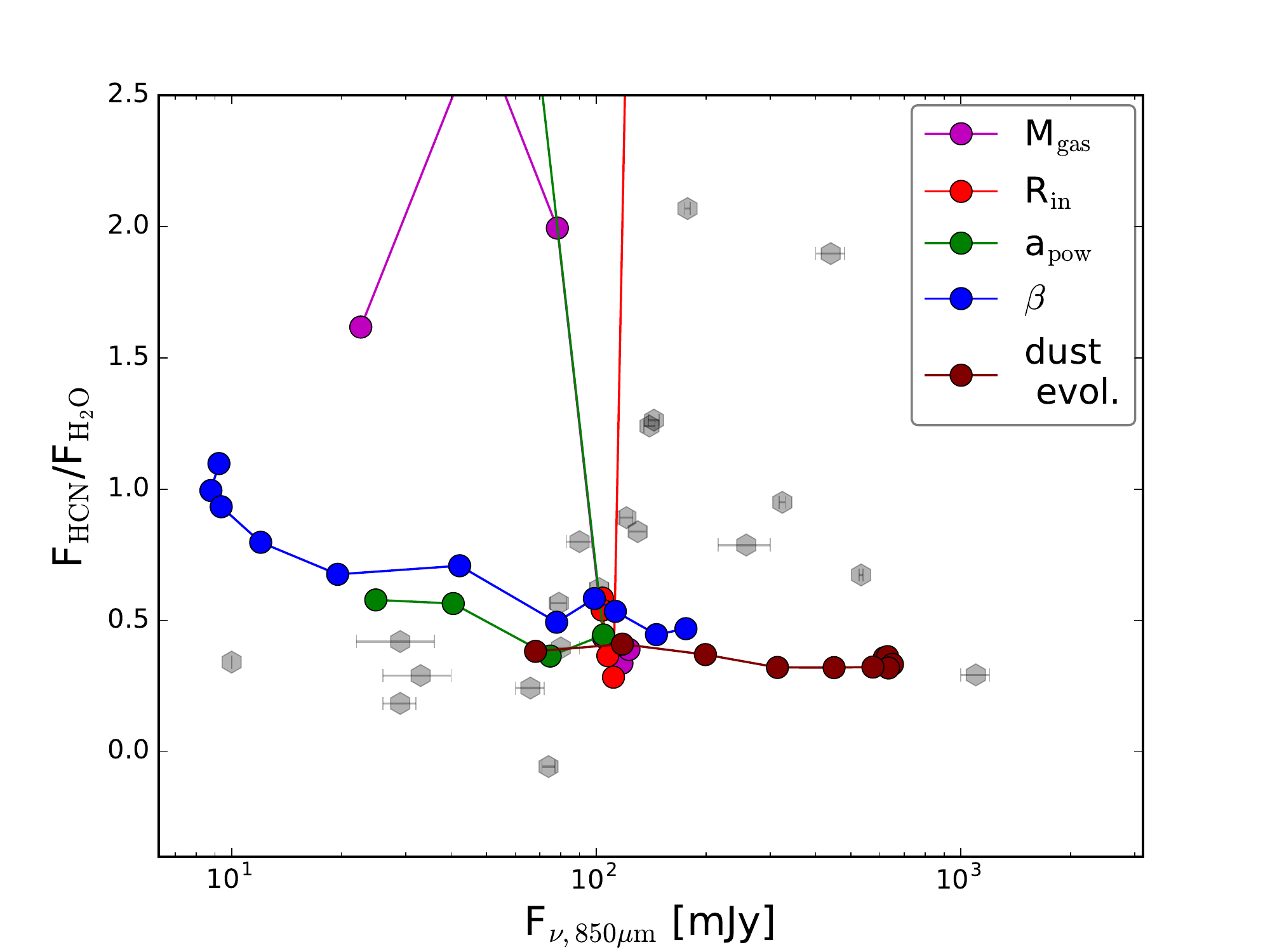}
\caption{HCN/H$_2$O MIR blend flux ratios versus 850~$\mu$m continuum flux from \citet{andrews2}, with indicated error bars. Observations and data collected for the objects (grey hexagons) are compared with the same quantities computed for our model series: disk gas mass $M_\mathrm{gas}$ (magenta), inner disk radius $R_\mathrm{in}$ (red), power law index of the grain size distribution $a_\mathrm{pow}$ (green), flaring angle $\beta$ (blue), and dust evolution models (brown) (symbols are the same as in previous figures). The red and green series extend beyond the shown range and include $a_\mathrm{pow}\,>\,3.5$ and $R_\mathrm{in}\,>\,5$~au.} 
\label{fig:mass}
\end{figure}

Figure~\ref{fig:mass} shows that 
the model series with a constant disk dust mass (blue, green) still span about a factor of 10 in the sub-mm flux.
We know from many previous work that the 
inference of dust mass from sub-mm fluxes can be affected by many other disk properties such as the flaring index, which changes the disk averaged dust temperature \citep[see also][]{woitke4}. The model series varying the power law index of the dust grain size distribution (green) illustrates how dust opacity affects the sub-mm fluxes by a factor of $\sim\!5$. Disks with a small flaring angle (blue) experience lower irradiation from the central star, making the disk  cooler overall and, hence, appearing fainter in the sub-mm.

In the dust evolution model, there are changes in both the dust opacities and the disk mass. The larger grains settle and drift inward efficiently; this changes the dust size distribution locally and so the dust opacity over time. Due to viscous accretion ($\alpha_\mathrm{vis}\!=\!10^{-3}$), the disk mass decreases with time by about two orders of magnitude between 10$^4$ and 10$^7$\!~yr. Figure~\ref{fig:mass} shows that the dust evolution model series (brown) evolves towards lower $850$~$\mu$m fluxes, while the HCN/H$_2$O blend ratio stays fairly constant, or slightly decreases. This result is rather opposite to what has been found by \citet{najita1}. Our blend ratios in this series agree well with the lower end of the range of observed values ($\sim\,0.3$). 

Most of the model series discussed above change one parameter at a time. We note that the grain power law, the flaring angle and the dust evolution series show moderate changes in the HCN/H$_2$O blend ratio, while changing the sub-mm flux substantially. However, some parameters show an orthogonal behaviour, they change the blend ratio substantially, while affecting the sub-mm flux to a lesser extent. For example, disk models that are highly truncated ($R_\mathrm{in}\!>\!1~$au red) can bring the HCN/H$_2$O blend ratio up to values in excess of 10$^2$. This is due to H$_2$O MIR blend fluxes decreasing below 10$^{-18}$~W/m$^2$ (Fig.~\ref{fig:model}) because they can no longer be excited. Such low fluxes would not, however, have been detectable in Spitzer IRS spectra. Disk models with very low flaring angles ($\beta\!\ll\!1$) also produce HCN/H$_2$O blend ratios higher than 1. Finally, also increasing the gas mass (without changing the dust mass, magenta series) strongly affects the HCN/H$_2$O blend ratios, with the ratio strongly increasing in line with the decrease in $M_{\rm gas}$. Thus, to conclude, varying multiple disk parameters together can easily span the observed range of HCN/H$_2$O blend ratios versus sub-mm fluxes (grey hexagons in Fig.~\ref{fig:mass}). Additional observables would be required to break these degeneracies.

\subsection{Outer radius and the HCN/H$_2$O ratio}

\begin{figure}
\includegraphics[width=0.5\textwidth]{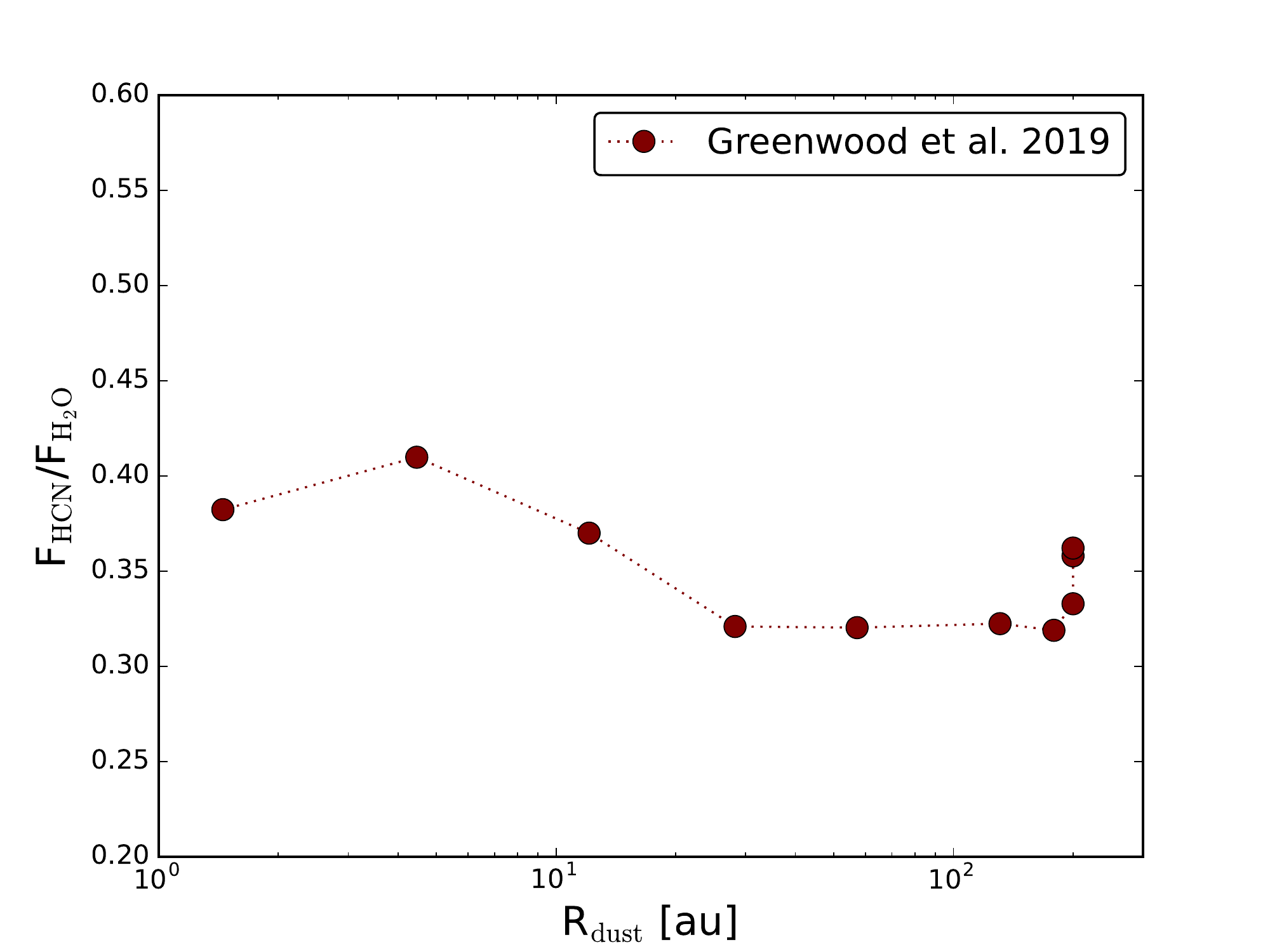}
\caption{HCN/H$_2$O MIR blend flux ratios versus estimated disk dust outer radius for the dust evolution model series; the evolution proceeds over 10~Myr.} 
\label{fig:migraine}
\end{figure}

\citet{bp2020} found the H$_2$O/HCN blend ratio to be anti-correlated to the dust outer radius, as measured from the sub-mm continuum emission. In our dust evolution model series, the disk size naturally evolves since the implemented dust evolution causes grains to grow and to drift inwards, thus affecting the local dust-to-gas mass ratio. This in turn decreases the surface density with time \citep[for detailed figures illustrating this, see][]{greenwood}. We defined a critical dust column density, $N_{\rm cr}$, in the initial disk set-up ($t=0$) at two times the tapering radius, $R_{\rm tap}$. We consider this value of $N_{\rm cr}$ as a proxy for the evolution of the 'observable' disk size, as seen in the ALMA continuum images. Extracting the location of $N_{\rm cr}$ from the time evolution series shows indeed a shrinking dust disk size as dust evolution proceeds (Fig.~\ref{fig:migraine}).

However, the HCN/H$_2$O blend ratio stays relatively constant, changing within 20\%, as already expected from the result in Fig.~\ref{fig:mass}. 
Within the molecular blend-emitting region, the dust evolution models produce a drop of 2~dex in the dust-to-gas ratio and 4~dex in the dust column density in the upper disk layers over the entire time evolution ($\sim$10$^5$-10$^7$~yr), while the gas surface density decreases by only a factor of 2 \citep{greenwood1}. This change affects both molecular emission bands (HCN and H$_2$O) in a similar manner.

 A single model series again cannot independently describe the full observed dynamic range in blend ratio versus disk dust size. In addition, we would need to combine multiple parameters that change the disk structure to explain the observed trends, namely,\ a parameter that affects the HCN/H$_2$O blend ratio combined with dust evolution. We expand on this issue in the next section.

\section{Discussion}\label{sec:dis}\vspace{5mm}

In \citet{antonellini2}, it was shown that the MIR H$_2$O emission is significantly affected by dust opacity effects, such as the dust content and/or the grain size, but independently from the dust composition. Disks where the water flux is strong in the Spitzer regime could have a reduced inner disk opacity due to dust depletion, different dust sizes, very strong settling, and, in a word, dust processing. The opposite effect would be produced by gas depletion in the inner disk regions (within 10~au). Such an effect can be the result of winds launched from the innermost part of the disk if they do not efficiently entrain dust grains. 

The dust evolution models considered here do not consider the effect of ice radial drift and so, they ignore the enrichment of the gas reservoir due to this process, as predicted by \citet{Ciesla2006,Krijt2020}. However, for MIR blends that are optically thick, any further inward gas enrichment (from ice transport and sublimation) should not affect the blend flux.

A key question to be considered at this point is whether it is the gas or the dust that evolves first. Analysing how our gas mass series changes the water, along with the HCN and spectral slope, we find unexpected trends. Gas-depleted disks are expected to have bluer MIR SEDs with respect to a gas enriched model or a standard $M_{\rm gas}$ value. Parameters such as the dust size power-law distribution and (in a more complete picture) the dust evolution, on the contrary, produce a trend where the MIR H$_2$O flux increases with the dust evolving along the disk lifetime. This suggests that disks are born opaque and with weak MIR blend features. During the dust evolution the object then experiences a phase where the MIR blend flux is boosted by the opacity overall decreasing due to grain growth and settling proceeding over time.

What we see in the correlations between MIR HCN, H$_2$O, and SEDs slopes is not an indication of disks with different gas mass, but rather a snapshot of a population of objects in different evolutionary phases. This is further supported by the fact that (at the level of precision required here), the 850~$\mu$m flux alone cannot serve as a direct disk mass tracer.

Recently, \citet{li1} investigated through hydrodynamical simulations how the presence of dust rings in disks affects the local grain size distribution via trapping and hence also the mm slope. The study found that the mm continuum flux and the slope anti-correlate due to dust growth and continuum optical depth. Such a scenario has not been explored by our 'monolithic' disk models, but should be analysed in follow-up works. Therefore, more investigation on the connection between the behaviour of MIR blends and dust continuum emission is needed in order to be able to use their diagnostic power for protoplanetary disk properties. The recently launched James Webb Space Telescope (JWST) is poised to collect data for tens of disks, many of which have existing spatially resolved ALMA data. Hence, much more detailed follow-up studies will become soon possible.

Simplified modelling of parametrised disk series and considerations based on a small set of observables such as those presented here still provide important insights into the effects driving observations. In \citet{antonellini,antonellini1,antonellini3}, the single-parameter investigation allowed to highlight the most influencial disk properties in determining MIR and far-infrared (FIR) H$_2$O spectra, as well as CO ro-vibrational transitions. We have proven this focussed approach to be more efficient than running an entire grid of complex models with a wide combination of properties. Single-parameter modelling allows for  elements of the full picture of disk structure and evolution to be built  and serve as a testing ground for  new data that will become available from JWST.  


\section{Conclusions}\label{sec:con}\vspace{5mm}

 In this paper, we focus on three main observables: MIR molecular blend fluxes, SED spectral indices, and continuum flux in the sub-mm. Their dynamic range and variations have been explained in previous works with different disk evolutionary stages and scenarios. Here, we use existing published thermo-chemical disk model series to revisit those interpretations.
\indent

Observations from \citet{najita1} can be explained with dust evolution, where a modest variation of the dust MIR opacity can boost the blend fluxes in this regime by more than a dex (a factor of $\sim$~20 for HCN and $\sim$~50 for H$_2$O). Dust evolution produces both the observed absolute fluxes of both these molecules as well as their correlation. Models show that a change in disk inner radius affects only the HCN flux and not water, producing potentially high blend ratios for these two species. 

Disk models show that the MIR continuum slope is sensitive to a change in the disk structure and so, the flaring angle $\beta$, and the inner radius, $R_\mathrm{in}$, are the main parameters driving the observed large dynamic range. Moderate variations, in the range from -0.1 to 0.5 are consistent with our dust evolution model series.

The estimation of the disk gas mass from the 850\,$\mu$m continuum flux in our models has proven to be highly degenerate. For example, the flaring angle affecting the average disk temperature confounds simple submm mass estimates. For our dust evolution model series spanning 4~dex of mass evolution, the simple submm mass estimate only produces 0.5~dex variation. The dust-evolution model series shows a clear systematic decrease in the dust disk outer radius over time.

Based on observations, MIR H$_2$O blend fluxes and the n$_{13-25}$ spectral index show an anti-correlation and the HCN/H$_2$O flux ratio is anticorrelated to the submm disk mass and dust outer radius. This trend is consistent with the dust evolution as predicted in our disk modelling series. The dust growth and migration produce a lower opacity in the inner 10~au and, consequently, enhance the molecular blend fluxes in the Spitzer spectral regime as well as the MIR spectral index and the dust outer radius. However, not all aspects of the observed trends are captured equally well. Gas in the disk may evolve and get depleted, producing an inner cavity whose size increases over time -- a process that is not captured in the dust evolution series. Our models show that such a scenario is consistent with the Spitzer upper limits for the water blends in disks with red MIR SEDs. The higher sensitivity of JWST/MIRI will allow for the detection of a larger number of MIR molecular blends in disks, to disentangle line blending as well as weaker emission, thus allowing to potentially test our conclusions. This study demonstrates the need for a holistic approach both in observations and modelling, capturing the intricate interplay between gas and dust during the disk evolution.


\begin{acknowledgement}
Astrophysics at Queen's University Belfast is supported by a grant from the STFC (ST/P000312/1). IK acknowledges funding from the European Union H2020-MSCA-ITN-2019 under Grant Agreement no.\ 860470 (CHAMELEON). This research was also supported and inspired by discussions through the ISSI (International Space Science Institute) International Team collaboration ’Zooming In On Rocky Planet Formation’ (team 482).
\end{acknowledgement}

\bibliographystyle{aa}
\bibliography{DraftAnd}

\begin{thebibliography}{}
\expandafter\ifx\csname natexlab\endcsname\relax\def\natexlab#1{#1}\fi
\providecommand{\url}[1]{\href{#1}{#1}}

\bibitem[{{Andrews} \& {Williams}(2005)}]{andrews2}
{Andrews}, S.~M., \& {Williams}, J.~P. 2005, \apj, 631, 1134

\bibitem[{{Antonellini} {et~al.}(2020){Antonellini}, {Banzatti}, {Kamp}, {Thi},
  \& {Woitke}}]{antonellini3}
{Antonellini}, S., {Banzatti}, A., {Kamp}, I., {Thi}, W.~F., \& {Woitke}, P.
  2020, \aap, 637, A29

\bibitem[{{Antonellini} {et~al.}(2015){Antonellini}, {Kamp},
  {Riviere-Marichalar}, {Meijerink}, {Woitke}, {Thi}, {Spaans}, {Aresu}, \&
  {Lee}}]{antonellini}
{Antonellini}, S., {Kamp}, I., {Riviere-Marichalar}, P., {et~al.} 2015, \aap,
  582, A105

\bibitem[{{Antonellini} {et~al.}(2016){Antonellini}, {Kamp}, {Lahuis},
  {Woitke}, {Thi}, {Meijerink}, {Aresu}, {Spaans}, {G{\"u}del}, \&
  {Liebhart}}]{antonellini1}
{Antonellini}, S., {Kamp}, I., {Lahuis}, F., {et~al.} 2016, \aap, 585, A61

\bibitem[{{Antonellini} {et~al.}(2017){Antonellini}, {Bremer}, {Kamp},
  {Riviere-Marichalar}, {Lahuis}, {Thi}, {Woitke}, {Meijerink}, {Aresu}, \&
  {Spaans}}]{antonellini2}
{Antonellini}, S., {Bremer}, J., {Kamp}, I., {et~al.} 2017, \aap, 597, A72

\bibitem[{{Banzatti} {et~al.}(2012){Banzatti}, {Meyer}, {Bruderer}, {Geers},
  {Pascucci}, {Lahuis}, {Juh{\'a}sz}, {Henning}, \& {{\'A}brah{\'a}m}}]{banz12}
{Banzatti}, A., {Meyer}, M.~R., {Bruderer}, S., {et~al.} 2012, \apj, 745, 90

\bibitem[{{Banzatti} {et~al.}(2020){Banzatti}, {Pascucci}, {Bosman}, {Pinilla},
  {Salyk}, {Herczeg}, {Pontoppidan}, {Vazquez}, {Watkins}, {Krijt}, {Hendler},
  \& {Long}}]{bp2020}
{Banzatti}, A., {Pascucci}, I., {Bosman}, A.~D., {et~al.} 2020, \apj, 903, 124

\bibitem[{{Birnstiel} {et~al.}(2012){Birnstiel}, {Klahr}, \&
  {Ercolano}}]{birnstiel2012}
{Birnstiel}, T., {Klahr}, H., \& {Ercolano}, B. 2012, \aap, 539, A148

\bibitem[{{Birnstiel} {et~al.}(2011){Birnstiel}, {Ormel}, \&
  {Dullemond}}]{birnstiel}
{Birnstiel}, T., {Ormel}, C.~W., \& {Dullemond}, C.~P. 2011, \aap, 525, A11

\bibitem[{{Bosman} {et~al.}(2018){Bosman}, {Tielens}, \& {van
  Dishoeck}}]{bosman}
{Bosman}, A.~D., {Tielens}, A. G.~G.~M., \& {van Dishoeck}, E.~F. 2018, \aap,
  611, A80

\bibitem[{{Bruderer} {et~al.}(2015){Bruderer}, {Harsono}, \& {van
  Dishoeck}}]{Bruderer2015}
{Bruderer}, S., {Harsono}, D., \& {van Dishoeck}, E.~F. 2015, \aap, 575, A94

\bibitem[{{Carr} \& {Najita}(2008)}]{carr}
{Carr}, J.~S., \& {Najita}, J.~R. 2008, Science, 319, 1504

\bibitem[{{Carr} \& {Najita}(2011)}]{carr1}
---. 2011, \apj, 733, 102

\bibitem[{{Ciesla} \& {Cuzzi}(2006)}]{Ciesla2006}
{Ciesla}, F.~J., \& {Cuzzi}, J.~N. 2006, \icarus, 181, 178

\bibitem[{{Draine}(2003)}]{draine}
{Draine}, B.~T. 2003, \araa, 41, 241

\bibitem[{{Dubrulle} {et~al.}(1995){Dubrulle}, {Morfill}, \&
  {Sterzik}}]{dubrulle}
{Dubrulle}, B., {Morfill}, G., \& {Sterzik}, M. 1995, \icarus, 114, 237

\bibitem[{{Evans} {et~al.}(2009{\natexlab{a}}){Evans}, {Calvet}, {Cieza},
  {Forbrich}, {Hillenbrand}, {Lada}, {Mer{\'{\i}}n}, {Strom}, \&
  {Watson}}]{evans1}
{Evans}, N., {Calvet}, N., {Cieza}, L., {et~al.} 2009{\natexlab{a}}, ArXiv
  e-prints, arXiv:0901.1691

\bibitem[{{Evans} {et~al.}(2009{\natexlab{b}}){Evans}, {Dunham},
  {J{\o}rgensen}, {Enoch}, {Mer{\'{\i}}n}, {van Dishoeck}, {Alcal{\'a}},
  {Myers}, {Stapelfeldt}, {Huard}, {Allen}, {Harvey}, {van Kempen}, {Blake},
  {Koerner}, {Mundy}, {Padgett}, \& {Sargent}}]{evans}
{Evans}, II, N.~J., {Dunham}, M.~M., {J{\o}rgensen}, J.~K., {et~al.}
  2009{\natexlab{b}}, \apjs, 181, 321

\bibitem[{{Furlan} {et~al.}(2006){Furlan}, {Hartmann}, {Calvet}, {D'Alessio},
  {Franco-Hern{\'a}ndez}, {Forrest}, {Watson}, {Uchida}, {Sargent}, {Green},
  {Keller}, \& {Herter}}]{furlan2}
{Furlan}, E., {Hartmann}, L., {Calvet}, N., {et~al.} 2006, \apjs, 165, 568

\bibitem[{{Furlan} {et~al.}(2009){Furlan}, {Watson}, {McClure}, {Manoj},
  {Espaillat}, {D'Alessio}, {Calvet}, {Kim}, {Sargent}, {Forrest}, \&
  {Hartmann}}]{furlan1}
{Furlan}, E., {Watson}, D.~M., {McClure}, M.~K., {et~al.} 2009, \apj, 703, 1964

\bibitem[{{Greenwood} {et~al.}(2019{\natexlab{a}}){Greenwood}, {Kamp},
  {Waters}, {Woitke}, \& {Thi}}]{greenwood}
{Greenwood}, A.~J., {Kamp}, I., {Waters}, L.~B.~F.~M., {Woitke}, P., \& {Thi},
  W.~F. 2019{\natexlab{a}}, \aap, 626, A6

\bibitem[{{Greenwood} {et~al.}(2019{\natexlab{b}}){Greenwood}, {Kamp},
  {Waters}, {Woitke}, \& {Thi}}]{greenwood1}
---. 2019{\natexlab{b}}, \aap, 631, A81

\bibitem[{{Houck} {et~al.}(2004){Houck}, {Roellig}, {van Cleve}, {Forrest},
  {Herter}, {Lawrence}, {Matthews}, {Reitsema}, {Soifer}, {Watson}, {Weedman},
  {Huisjen}, {Troeltzsch}, {Barry}, {Bernard-Salas}, {Blacken}, {Brandl},
  {Charmandaris}, {Devost}, {Gull}, {Hall}, {Henderson}, {Higdon}, {Pirger},
  {Schoenwald}, {Sloan}, {Uchida}, {Appleton}, {Armus}, {Burgdorf},
  {Fajardo-Acosta}, {Grillmair}, {Ingalls}, {Morris}, \& {Teplitz}}]{houck}
{Houck}, J.~R., {Roellig}, T.~L., {van Cleve}, J., {et~al.} 2004, \apjs, 154,
  18

\bibitem[{{Kamp} {et~al.}(2010){Kamp}, {Tilling}, {Woitke}, {Thi}, \&
  {Hogerheijde}}]{kamp4}
{Kamp}, I., {Tilling}, I., {Woitke}, P., {Thi}, W.-F., \& {Hogerheijde}, M.
  2010, \aap, 510, A18

\bibitem[{{Krijt} {et~al.}(2020){Krijt}, {Bosman}, {Zhang}, {Schwarz},
  {Ciesla}, \& {Bergin}}]{Krijt2020}
{Krijt}, S., {Bosman}, A.~D., {Zhang}, K., {et~al.} 2020, \apj, 899, 134

\bibitem[{{Lada} {et~al.}(2006){Lada}, {Muench}, {Luhman}, {Allen}, {Hartmann},
  {Megeath}, {Myers}, {Fazio}, {Wood}, {Muzerolle}, {Rieke}, {Siegler}, \&
  {Young}}]{lada2}
{Lada}, C.~J., {Muench}, A.~A., {Luhman}, K.~L., {et~al.} 2006, \aj, 131, 1574

\bibitem[{{Li} {et~al.}(2019){Li}, {Li}, {Ricci}, {Li}, {Birnstiel}, {Isella},
  {Ansdell}, {Yuan}, {Dr{\k{a}}{\.z}kowska}, \& {Stammler}}]{li1}
{Li}, Y.-P., {Li}, H., {Ricci}, L., {et~al.} 2019, \apj, 878, 39

\bibitem[{{Meijerink} {et~al.}(2009){Meijerink}, {Pontoppidan}, {Blake},
  {Poelman}, \& {Dullemond}}]{Meijerink}
{Meijerink}, R., {Pontoppidan}, K.~M., {Blake}, G.~A., {Poelman}, D.~R., \&
  {Dullemond}, C.~P. 2009, \apj, 704, 1471

\bibitem[{{Mulders} \& {Dominik}(2012)}]{mulders1}
{Mulders}, G.~D., \& {Dominik}, C. 2012, \aap, 539, A9

\bibitem[{{Najita} {et~al.}(2013){Najita}, {Carr}, {Pontoppidan}, {Salyk}, {van
  Dishoeck}, \& {Blake}}]{najita1}
{Najita}, J.~R., {Carr}, J.~S., {Pontoppidan}, K.~M., {et~al.} 2013, \apj, 766,
  134

\bibitem[{{Pontoppidan} {et~al.}(2009){Pontoppidan}, {Meijerink}, {Dullemond},
  \& {Blake}}]{pontoppidan}
{Pontoppidan}, K.~M., {Meijerink}, R., {Dullemond}, C.~P., \& {Blake}, G.~A.
  2009, \apj, 704, 1482

\bibitem[{{Pontoppidan} {et~al.}(2010{\natexlab{a}}){Pontoppidan}, {Salyk},
  {Blake}, \& {K{\"a}ufl}}]{pontoppidan1}
{Pontoppidan}, K.~M., {Salyk}, C., {Blake}, G.~A., \& {K{\"a}ufl}, H.~U.
  2010{\natexlab{a}}, \apjl, 722, L173

\bibitem[{{Pontoppidan} {et~al.}(2010{\natexlab{b}}){Pontoppidan}, {Salyk},
  {Blake}, {Meijerink}, {Carr}, \& {Najita}}]{pontoppidan2}
{Pontoppidan}, K.~M., {Salyk}, C., {Blake}, G.~A., {et~al.} 2010{\natexlab{b}},
  \apj, 720, 887

\bibitem[{{Salyk} {et~al.}(2007){Salyk}, {Blake}, {Boogert}, \&
  {Brown}}]{salyk1}
{Salyk}, C., {Blake}, G.~A., {Boogert}, A.~C.~A., \& {Brown}, J.~M. 2007,
  \apjl, 655, L105

\bibitem[{{Salyk} {et~al.}(2008){Salyk}, {Pontoppidan}, {Blake}, {Lahuis}, {van
  Dishoeck}, \& {Evans}}]{salyk}
{Salyk}, C., {Pontoppidan}, K.~M., {Blake}, G.~A., {et~al.} 2008, \apjl, 676,
  L49

\bibitem[{{Salyk} {et~al.}(2011){Salyk}, {Pontoppidan}, {Blake}, {Najita}, \&
  {Carr}}]{salyk2}
{Salyk}, C., {Pontoppidan}, K.~M., {Blake}, G.~A., {Najita}, J.~R., \& {Carr},
  J.~S. 2011, \apj, 731, 130

\bibitem[{{Thi} {et~al.}(2011){Thi}, {Woitke}, \& {Kamp}}]{thi2011}
{Thi}, W.~F., {Woitke}, P., \& {Kamp}, I. 2011, \mnras, 412, 711

\bibitem[{{Woitke} {et~al.}(2009){Woitke}, {Kamp}, \& {Thi}}]{woitke}
{Woitke}, P., {Kamp}, I., \& {Thi}, W.-F. 2009, \aap, 501, 383

\bibitem[{{Woitke} {et~al.}(2018){Woitke}, {Min}, {Thi}, {Roberts}, {Carmona},
  {Kamp}, {M{\'e}nard}, \& {Pinte}}]{woitke6}
{Woitke}, P., {Min}, M., {Thi}, W.~F., {et~al.} 2018, \aap, 618, A57

\bibitem[{{Woitke} {et~al.}(2016){Woitke}, {Min}, {Pinte}, {Thi}, {Kamp},
  {Rab}, {Anthonioz}, {Antonellini}, {Baldovin-Saavedra}, {Carmona}, {Dominik},
  {Dionatos}, {Greaves}, {G{\"u}del}, {Ilee}, {Liebhart}, {M{\'e}nard},
  {Rigon}, {Waters}, {Aresu}, {Meijerink}, \& {Spaans}}]{woitke4}
{Woitke}, P., {Min}, M., {Pinte}, C., {et~al.} 2016, \aap, 586, A103

\bibitem[{{Zhang} {et~al.}(2013){Zhang}, {Pontoppidan}, {Salyk}, \&
  {Blake}}]{zhang}
{Zhang}, K., {Pontoppidan}, K.~M., {Salyk}, C., \& {Blake}, G.~A. 2013, \apj,
  766, 82

\end{thebibliography}

\appendix

\section{Model series properties}\label{Ap1}

Table~\ref{CentralStars} provides a summary of the main properties of the central star and physical properties of the disk modelling series used here.

\begin{table*}
\centering
\caption{Overview of the models parameters for \citet{antonellini} and \citet{greenwood}}
\begin{tabular}{p{4.5cm}p{2.5cm}p{2.5cm}p{2.5cm}}
\hline\hline
\multicolumn{4}{c}{Central star and radiation field parameters}\\
\hline
Parameter & Symbol & \citet{antonellini} &  \citet{greenwood1} \\ \hline
Photospheric temperature & $T_\mathrm{eff}$ [K] & 4400 & 4000 \\
Stellar mass & $M_\mathrm{*}$ [M$_\mathrm{\odot}$] & 0.8 & 0.7\\
Stellar luminosity & $L_\mathrm{*}$ [L$_\mathrm{\odot}$] & 0.7 & 1.0\\
FUV excess & $L_\mathrm{UV}$/$L_\mathrm{*}$ & 0.01 & 0.01\\
UV power-law exponent & $p_\mathrm{UV}$ & 0.2 & 1.3\\
X-ray luminosity & $L_\mathrm{X}$ [erg/s] & 10$^{30}$ & 10$^{30}$\\
X-ray minimum energy & $E_\mathrm{min,X}$ [keV] & 0.1 & -\\
X-ray temperature & $T_\mathrm{X}$ [K] & 10$^7$ & -\\ \hline\hline
\multicolumn{4}{c}{Disk parameters fixed in the
series}\\
\hline
Parameter & Symbol & \citet{antonellini} & \citet{greenwood}\\ \hline
Radial $\times$ vertical grid points & $N_\mathrm{xx}\times\! N_\mathrm{zz}$ & 70 $\times$ 70 & 240 $\times$ 160\\
Outer radius & $R_\mathrm{out}$ [au] & 300 & 2000\\
Minimum dust size & $a_\mathrm{min}$ [$\mu$m] & 0.05 & 0.1\\
Maximum dust size & $a_\mathrm{max}$ [mm] & 1 & 1900\\
Dust composition & - & Draine Astrosilicates$^{1}$ & 60\% Amorph. Mg$_{0.7}$Fe$_{0.3}$SiO$_{3}$, 15\% Amorph. Carbon, 25\% vacuum \\
Reference radius & $R_\mathrm{0}$ [au] & 10 & 1 \\
Scale height at reference radius & $H_\mathrm{0}$ [au] &
100 & 0.05012\\
Scale height power-law index & $\beta$ & 1.13 & 1.15\\
Tapering-off radius & $R_\mathrm{taper}$ [au] & 200 & 100\\
Chemical heating efficiency & - & 0.2 & -\\
Settling description & - & Dubrulle & MD12$^2$\\
Cosmic-ray ionization rate & $\zeta_\mathrm{CRs}$ [s$^{-1}$] &
1.7$\times10^{-17}$ & -\\
Distance & $d$ [pc] & 140 & 150\\
Turbulence viscosity coefficient & $\alpha_\mathrm{vis}$ & 0.05 & -\\
Disk inclination & [$\degr$] & 30 & 45\\
\hline
\end{tabular}
\tablefoot{$^1$ \citet{draine}, $^2$ \citet{mulders1}}
\label{CentralStars}
\end{table*}

\section{H$_2$O- and HCN-emitting regions}

\begin{figure*}[!htb]
\includegraphics[width=0.4\textwidth]{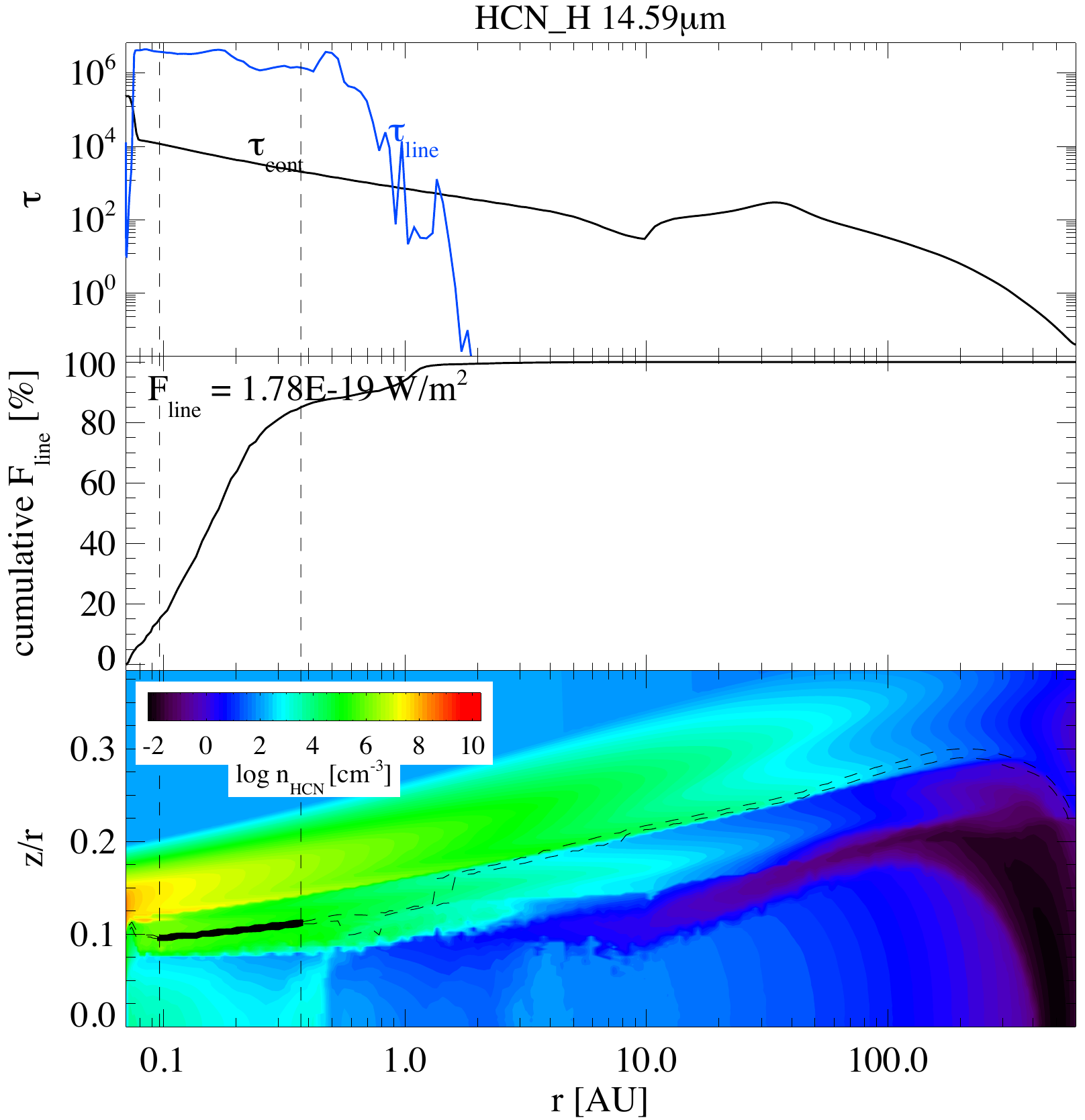} 
\includegraphics[width=0.41\textwidth]{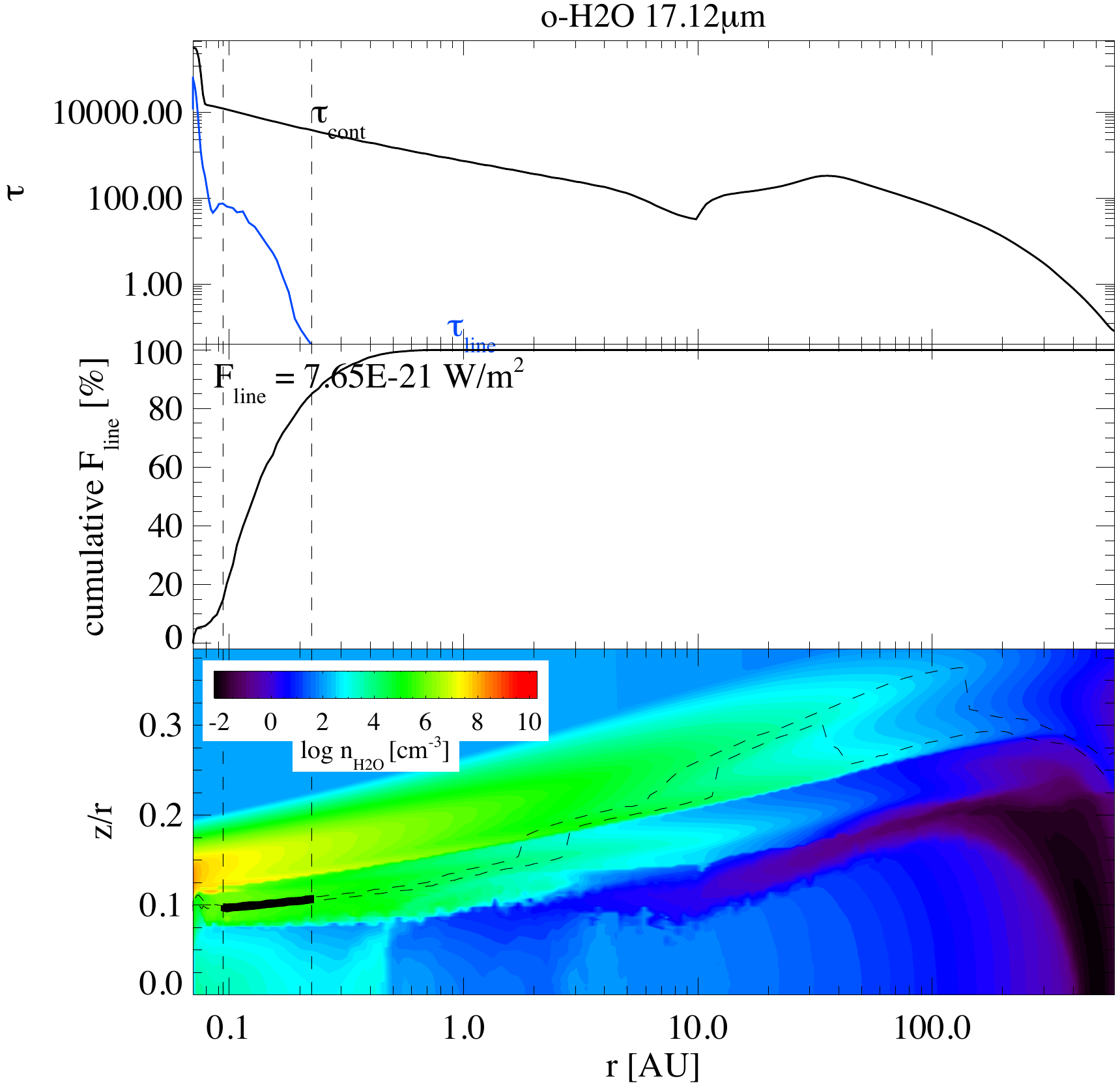}
\includegraphics[width=0.4\textwidth]{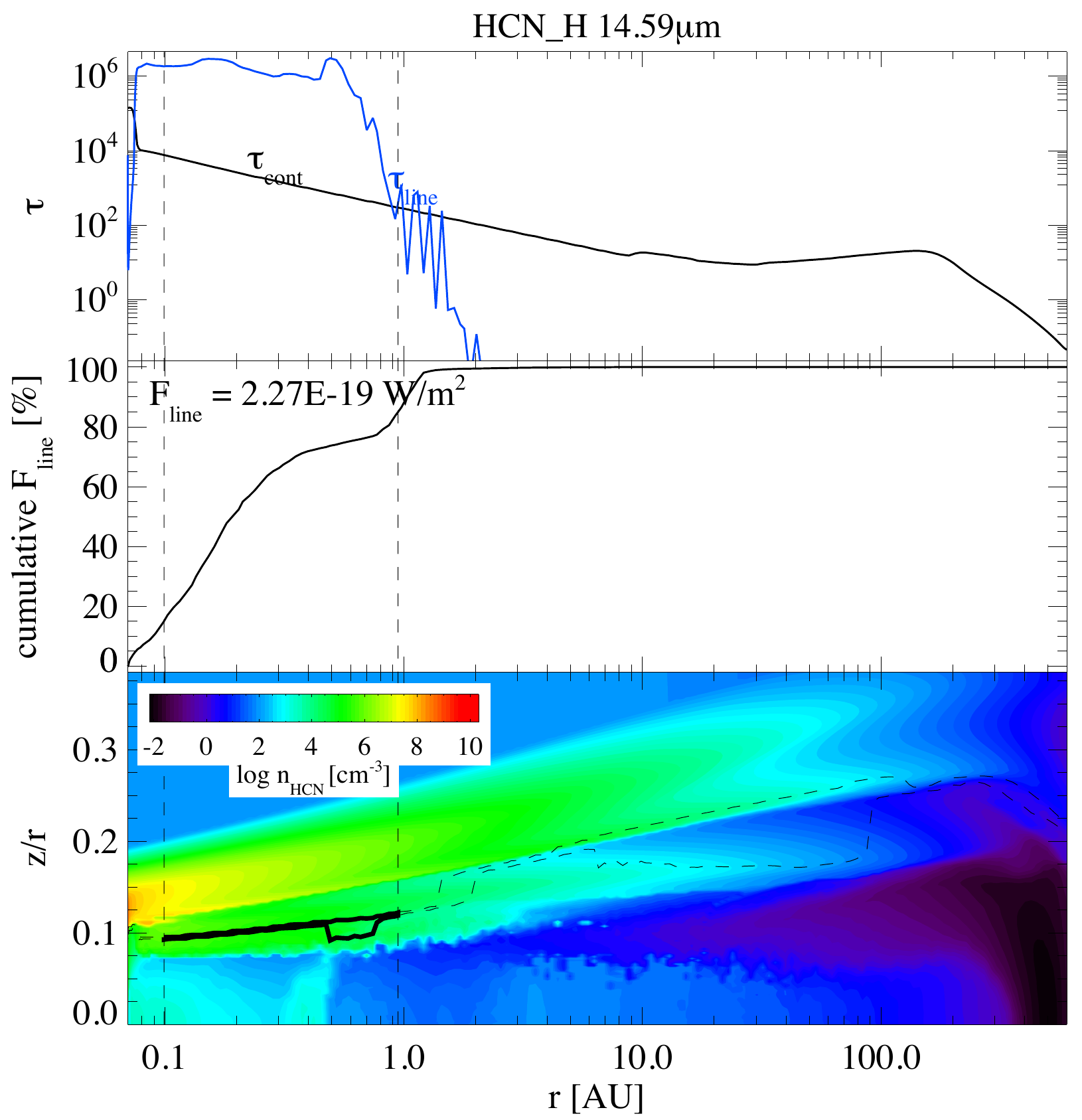} 
\includegraphics[width=0.4\textwidth]{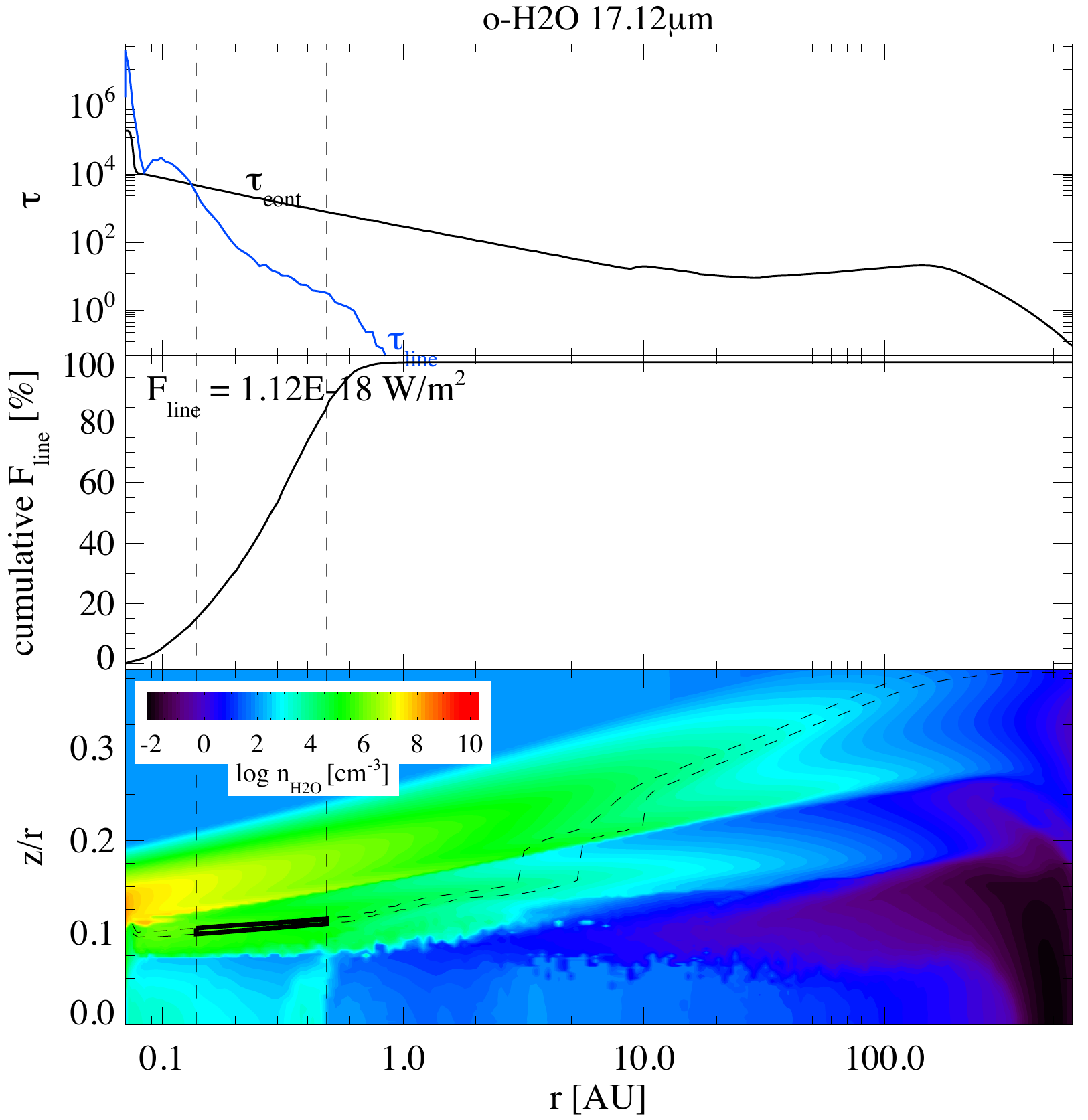}
\includegraphics[width=0.4\textwidth]{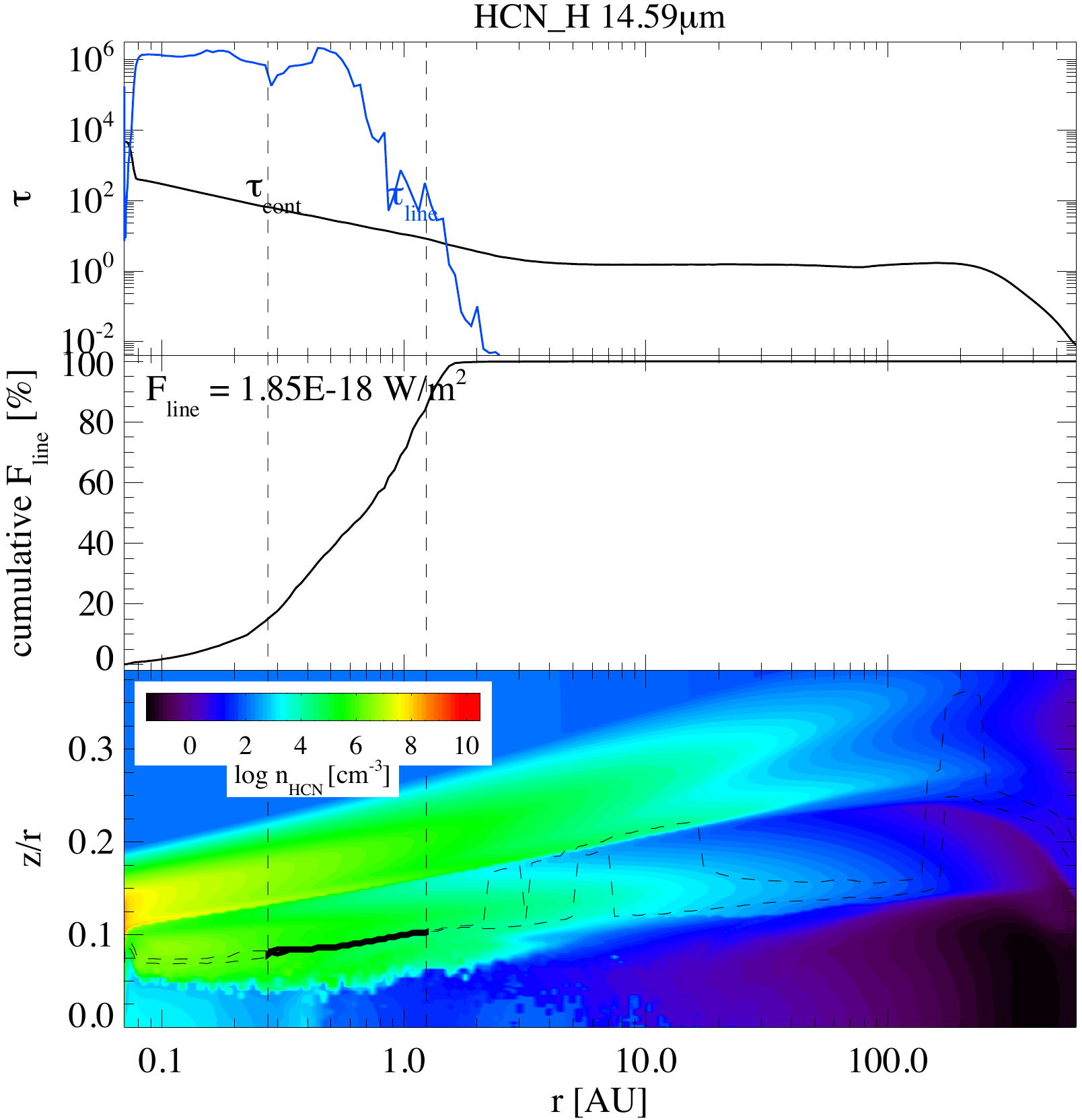}
\hspace{35mm}
\includegraphics[width=0.4\textwidth]{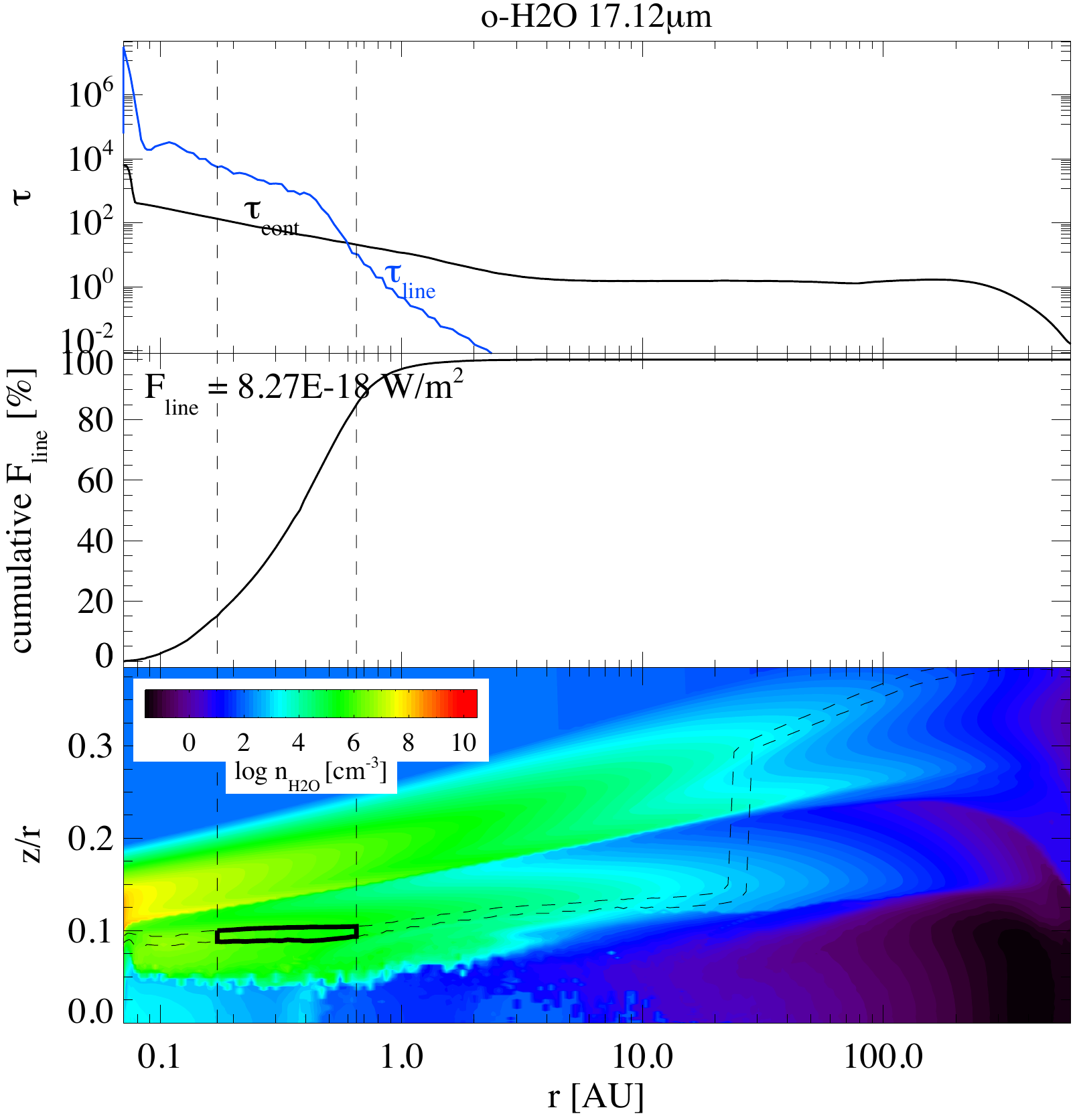}
\caption{Blend-emitting regions from evolutions models from \citet{greenwood} for o-H$_2$O blend at 17.12~$\mu$m and HCN blend at 14.59~$\mu$m with epochs 1.78$\cdot$10$^{4}$~yr (top), 1.78$\cdot$10$^{5}$~yr (middle), and 1.78$\cdot$10$^{6}$~yr (bottom). In each figure, the top panel shows the continuum optical depth (black line) and the blend optical depth (blue line). The middle panel shows the cumulative blend flux from vertical escape probability as a function of the radial distance from the star. The bottom panel shows the H$_2$O or HCN density in color scale with the region from which the vertical $\times$ radial integrated flux, amounting then to 49\% of the total blend flux, is emitted. This is reported as numbers in the middle slice of each plot (black contours, 15-85\% of the vertical and radial integrated blend flux, dashed lines).}
\label{EmReg}
\end{figure*}

Figure~\ref{EmReg} shows the emitting region for the HCN P(9) fundamental and o-H$_2$O v$_2$=1 8$_{3,6}$-7$_{4,3}$ hot band blends (black box) over-plotted on the species density, respectively. Here, the line and continuum opacity and the cumulative blend flux are also shown -- for three representative time steps of the dust evolution model series from \citet{greenwood}: 1.78$\cdot$10$^{4}$~yr, 1.78$\cdot$10$^{5}$~yr, and 1.78$\cdot$10$^{6}$~yr.

\section{Mid-IR spectral energy distributions and blend fluxes}

\begin{figure*}
\includegraphics[width=0.5\textwidth]{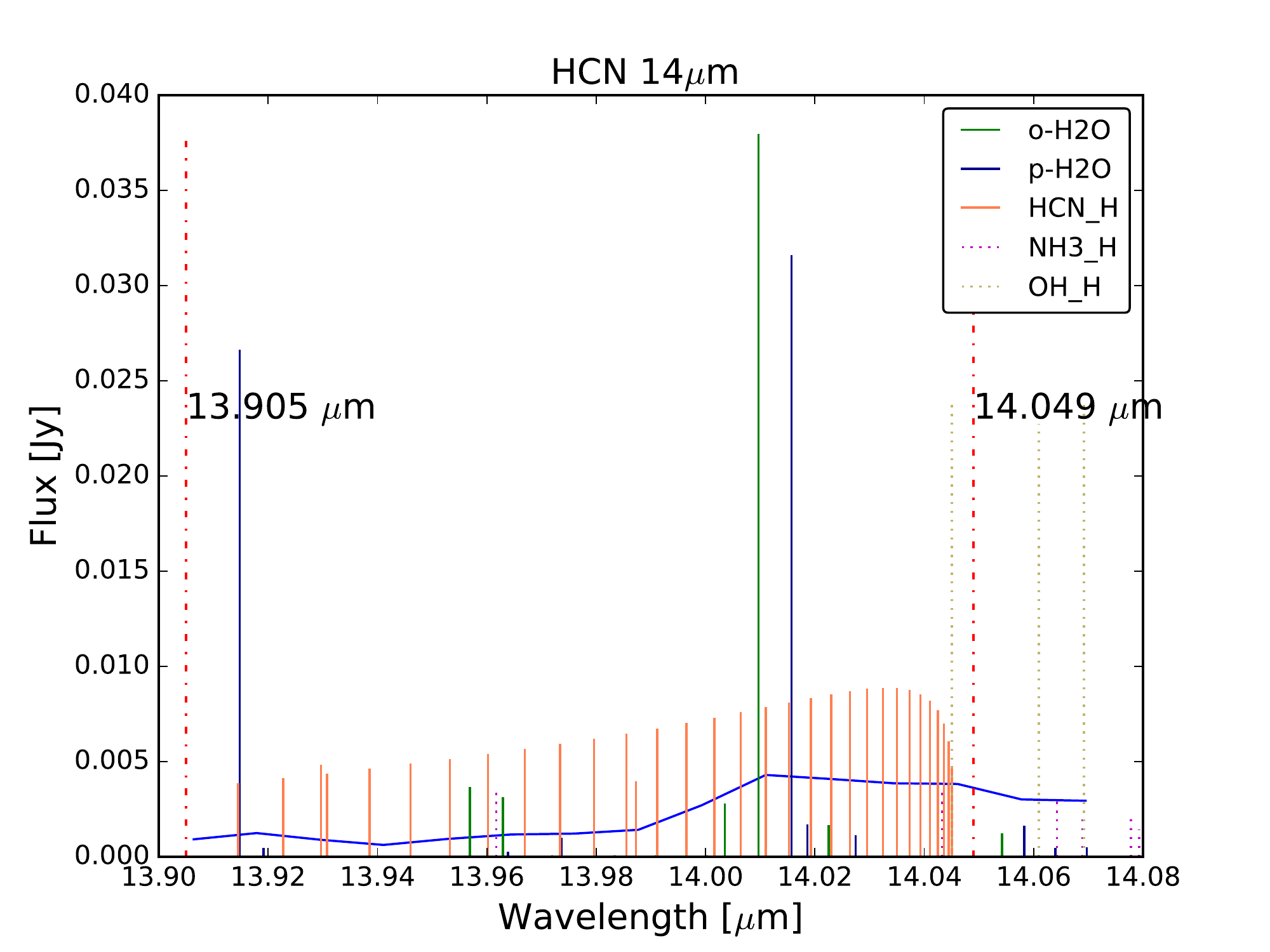}
\includegraphics[width=0.5\textwidth]{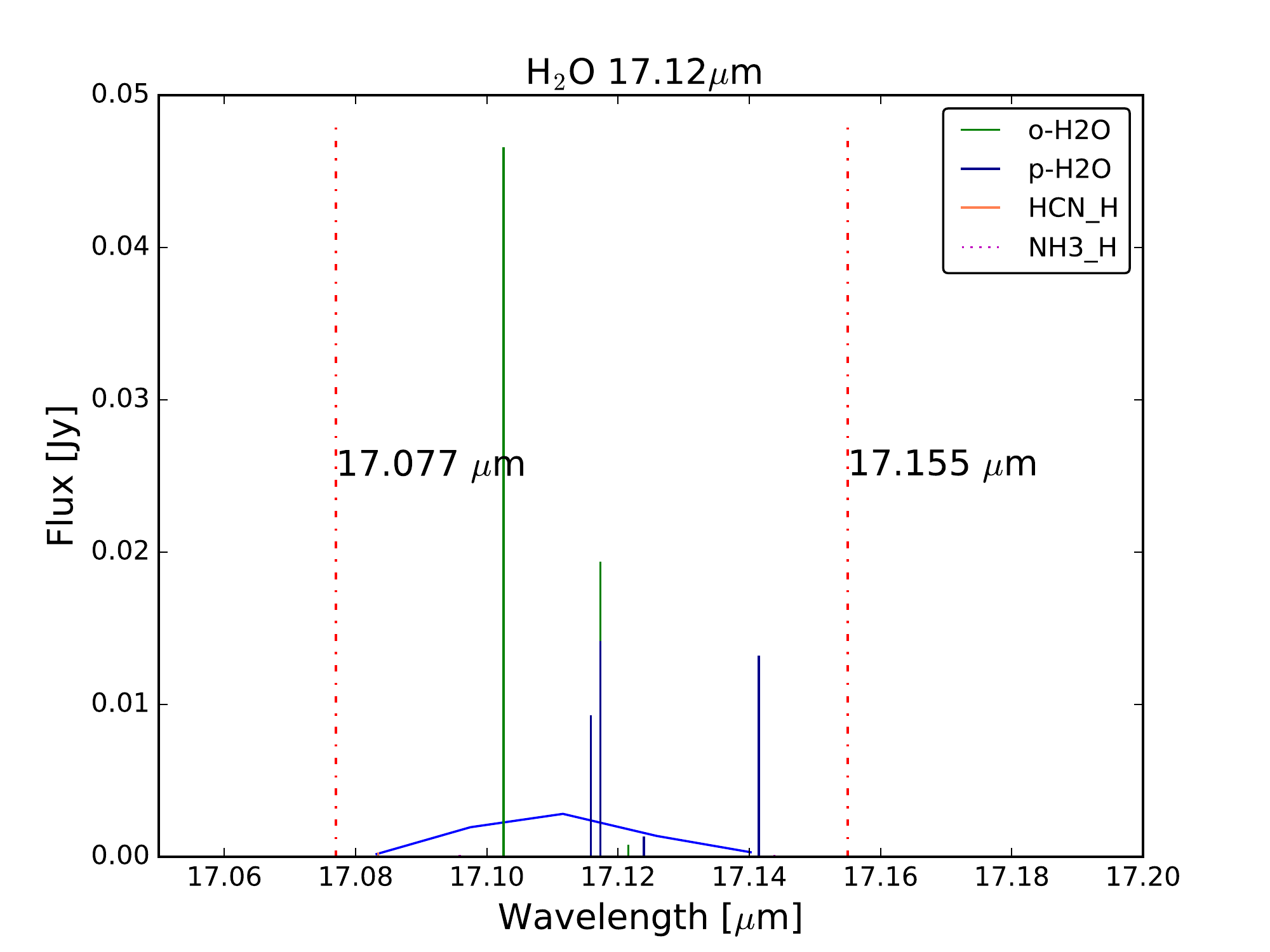}
\includegraphics[width=0.5\textwidth]{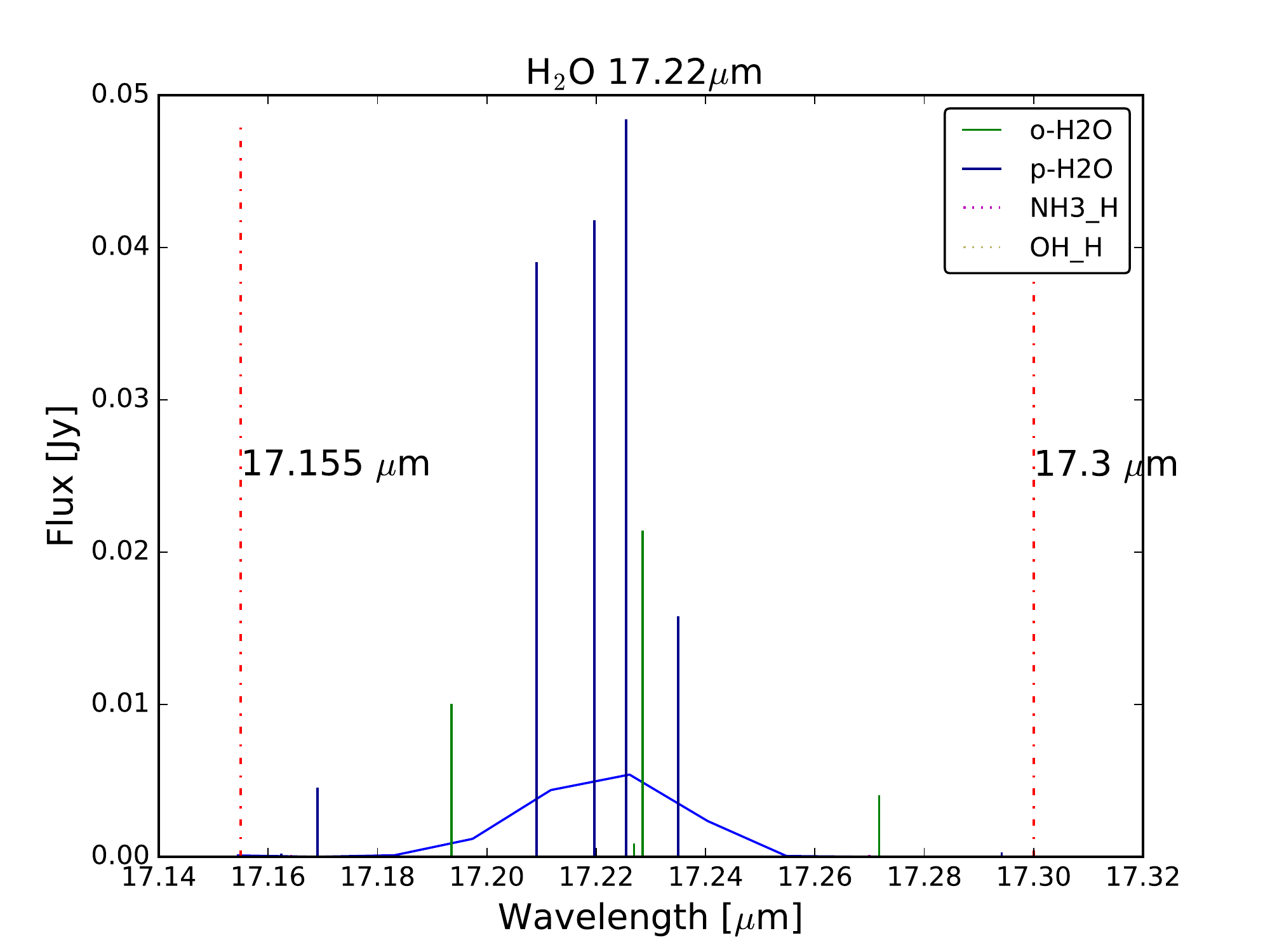}
\includegraphics[width=0.5\textwidth]{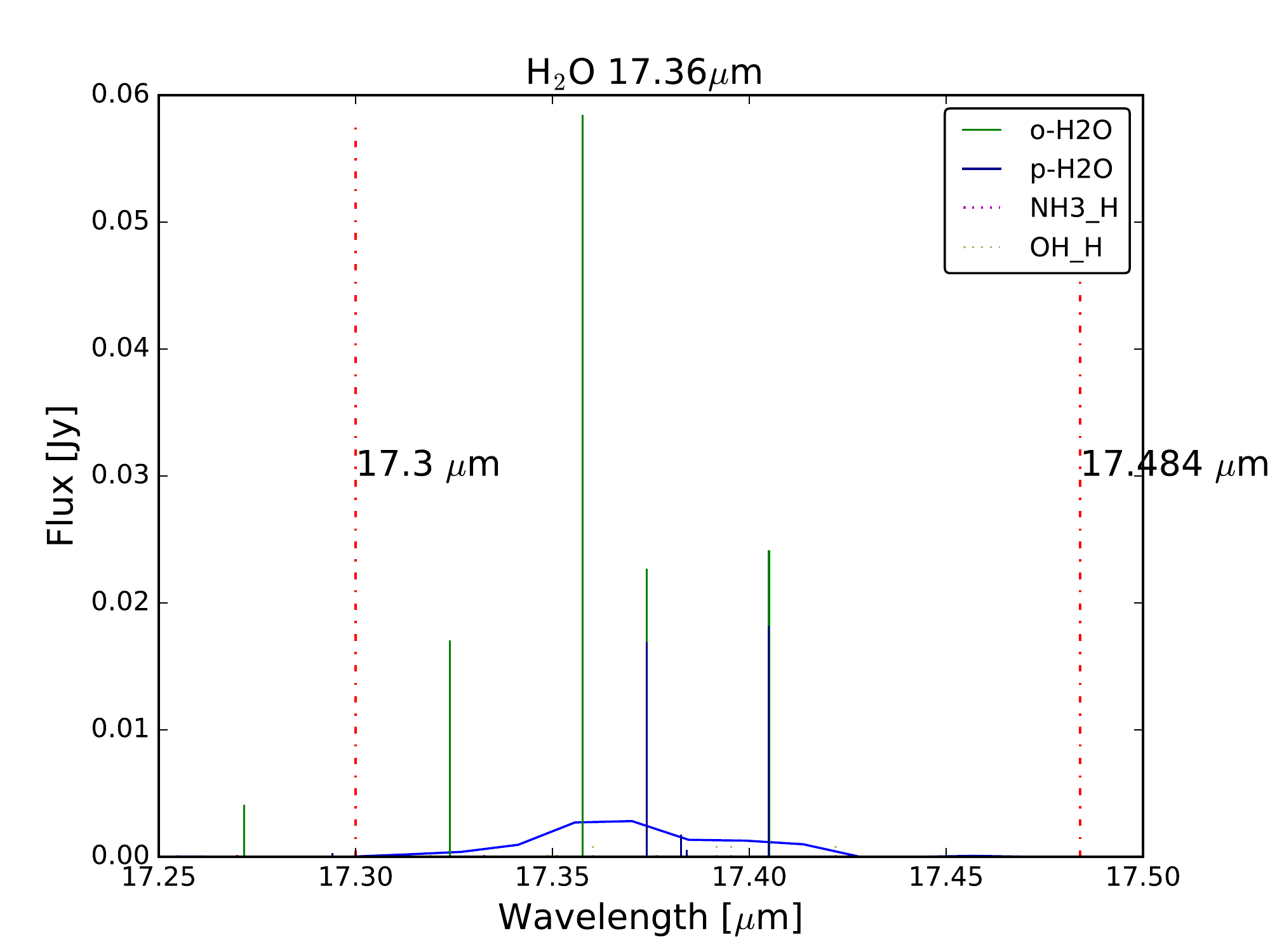}
\caption{Top left plot, Spitzer IRS regime for the 14~$\mu$m HCN blend. Top right, same plot for H$_2$O in the 17.12~$\mu$m regime. Bottom left, same plot for H$_2$O in the 17.22~$\mu$m regime. Bottom right, H$_2$O in the 17.36~$\mu$m regime. Each transition in the regime is plotted as a dark blue or green line (ortho and para H$_2$O), red blend for HCN ro-vibrational (LTE treatment, HITRAN database$^{(*)}$), dotted magenta line for NH$_3$ rovibrational (LTE treatment, HITRAN database$^{(*)}$), dotted khaki for OH (LTE treatment, HITRAN database$^{(*)}$). The blue solid curve represents the final blended IRS spectrum (R~=~600). Vertical dot-dashed lines delimit the range over which the blend flux is taken in \citet{najita1}.}
\label{600}
\end{figure*}

\begin{figure}
\includegraphics[width=0.5\textwidth]{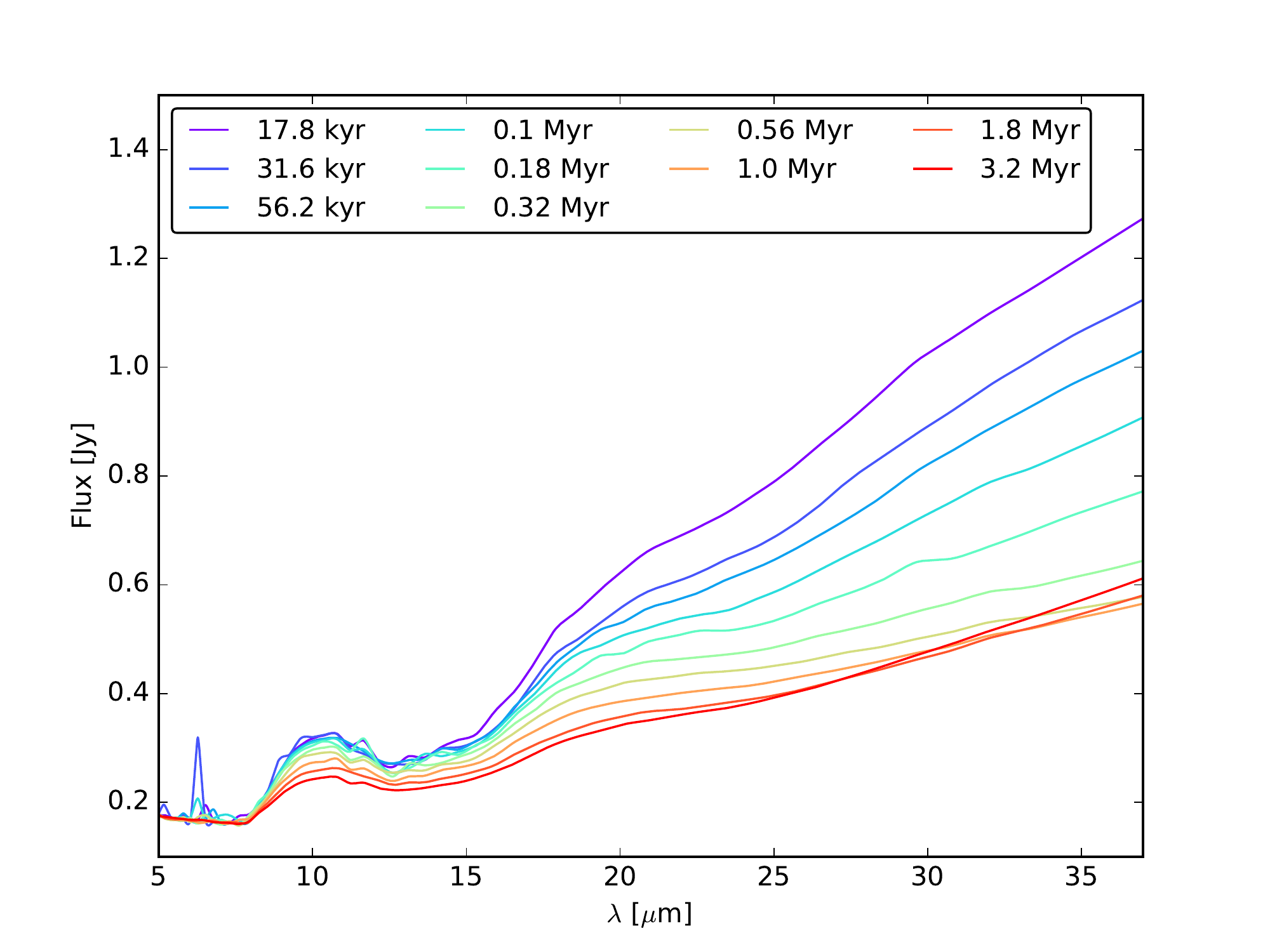}
\caption{SEDs in the range of the 10~$\mu$m \& 20~$\mu$m silicate emission features for the dust evolution model series from \citet{greenwood}. Molecular emission blends are not included to highlight the effects on the continuum.}
\label{Sil}
\end{figure}

We report here the details for extracting the blend fluxes from the MIR spectral regions of the models. 
Figure~\ref{600} shows the wavelength regions for the HCN and H$_2$O blends, the individual molecular blends in that region (shown by different colors) and also the wavelength boundaries used in the extraction procedure (vertical red dot-dashed lines). The data for this figure are taken from the standard disk model. Similarly, Fig.~\ref{Sil} shows the MIR SED in the wavelength range covered by the Spitzer IRS spectrograph for the dust evolution models. SEDs for the other modelling series can be found in \citet{antonellini}.

\footnotetext{(*) https://hitran.org}

\section{Normalised MIR water luminosity versus spectral index}

\begin{figure}
\includegraphics[width=0.5\textwidth]{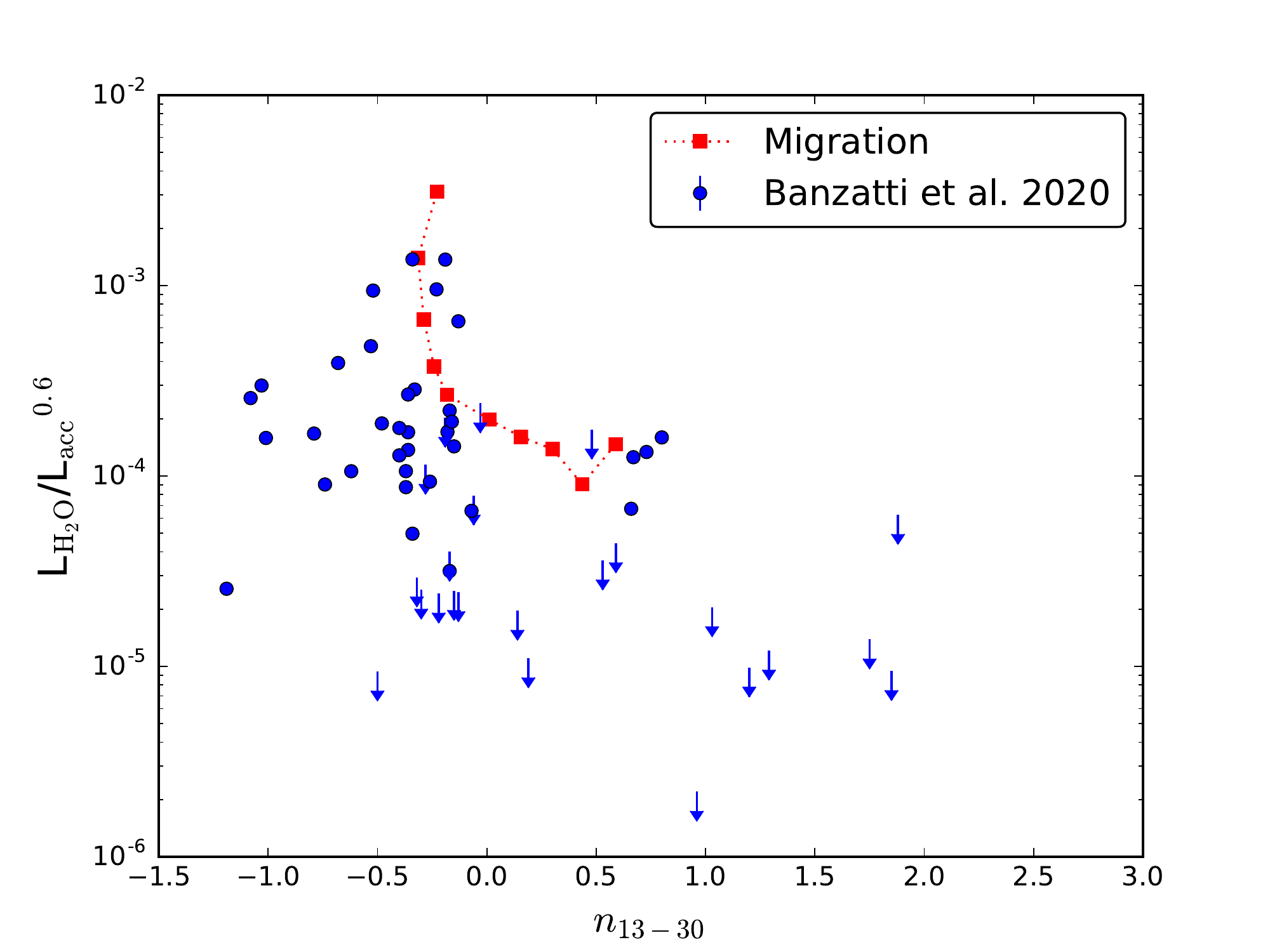}
\caption{Accretion-normalised MIR water luminosity versus the n$_{13-30}$ continuum spectral index. Blue dots and arrows (upper limits) are observations taken from Table~2 and 3 of \citet{bp2020}. Red squares are the values from the dust evolutionary models of \citet{greenwood}}
\label{Acc}
\end{figure}

In Fig.~\ref{Acc}, we show that the dust evolutionary models \citep{greenwood} are capable of reproducing the anti-correlation trend found in observations by \citet{bp2020}. The value of $L_{\rm accr}$ is estimated according to the formula in Eq.~\ref{Eqacc}, as follows:

\begin{equation}\label{Eqacc}
    L_\mathrm{acc} = L_\mathrm{*} \cdot f_\mathrm{UV} 
,\end{equation}

where $L_{\rm *}$ is the stellar luminosity, $f_{\rm UV}$ is the accretion contribution to the total stellar luminosity based on all the parameters provided as input to our models.

\end{document}